\documentclass[aps, prd, superscriptaddress, nofootinbib, preprintnumbers]{revtex4}
\usepackage{amsmath}
\usepackage[dvipdfmx,dvips]{graphicx}
\usepackage{color}

\newcommand{\lesssim}{\mathrel{\mathpalette\vereq<}}
\newcommand{\gtrsim}{\mathrel{\mathpalette\vereq>}}

\newcommand{\chushi}[1]{}

\newcommand{\dis}[1]{\begin{equation}\begin{split}#1\end{split}\end{equation}}
\newcommand{\eq}[1]{Eq.~(\ref{#1})}
\newcommand{\bfrac}[2]{{\left(\frac{#1}{#2} \right)  }}

\newcommand{\FTD}{F_{\rm TD}}
\newcommand{\MTD}{M_{\rm TD}}

\newcommand{\MP}{M_P}

\newcommand{\ev}{\,{\rm eV}}
\newcommand{\kev}{\,{\rm keV}}
\newcommand{\mev}{\,{\rm MeV}}
\newcommand{\gev}{\,{\rm GeV}}
\newcommand{\tev}{\,{\rm TeV}}

\begin{document}
\preprint{APCTP-Pre-2012-001, MISC-2012-01, PNUTP-12-A01  }

\title{%
  \hfill{\normalsize\vbox{%
  }}\\
  \vspace{-0.5cm}
  {\bf  Analysis of techni-dilaton as a dark matter candidate } }
\author{Ki-Young Choi}\thanks{
      {\tt kiyoung.choi@apctp.org}}
      \affiliation{Asia Pacific Center for Theoretical Physics,  POSTECH, Pohang 709-784, Korea.}
       \affiliation{Department of Physics, POSTECH, Pohang, Gyeongbuk 790-784, Republic of Korea.}
\author{Deog Ki Hong}\thanks{
      {\tt dkhong@pusan.ac.kr}}
      \affiliation{ Department of Physics,  
                    Pusan National University, Busan 609-735, Korea.}
\author{Shinya Matsuzaki}\thanks{
      {\tt synya@cc.kyoto-su.ac.jp}}
      \affiliation{ Maskawa Institute for Science and Culture, Kyoto Sangyo University, Motoyama, Kamigamo, Kita-Ku, Kyoto 603-8555, Japan}

\date{\today}

\begin{abstract} 
The almost conformal dynamics of walking technicolor (TC)   
implies the existence of the approximate scale invariance,  
which breaks down spontaneously by the condensation of anti-techni and techni-fermions. 
According to the Goldstone theorem, a spinless, parity-even particle, called  techni-dilaton (TD), 
then emerges at low energy. 
If TC exhibits an extreme walking,  TD  mass, $M_{\rm TD}$, is parametrically much smaller than 
that of techni-fermions ($\sim1\,{\rm TeV}$), while its decay constant is comparable to
the cutoff scale of walking TC.  
We analyze the light, decoupled TD as a dark matter candidate 
and study cosmological productions of TD, both thermal and non-thermal, 
in the early Universe. 
The thermal population is governed dominantly by single TD production processes involving vertices 
breaking the scale symmetry, while the non-thermal population is by the vacuum misalignment and is accumulated via 
harmonic and coherent oscillations of misaligned classical TD fields. 
The non-thermal population turns out to be dominant and large enough to explain the abundance of presently 
observed dark matter, 
while the thermal population is highly suppressed due to the large TD decay constant.    
Several cosmological and astrophysical limits on the light, decoupled TD are examined to find that  
the mass $M_{\rm TD}$ is constrained to be in a range, $0.01 {\rm eV} \lesssim M_{\rm TD} \lesssim 500$ eV.   
From the combined constraints on cosmological productions and astrophysical observations, 
we find that the light, decoupled TD can be a good dark matter candidate with the mass around a few hundreds of eV for typical models of (extreme) walking TC. We finally mention possible designated experiments to detect the TD dark matter. 

\end{abstract} 

\maketitle

\section{Introduction}

The standard model (SM) of particle physics has been extremely successful in describing all the interactions of elementary particles,  apart from the gravitational interaction. 
A candidate for Higgs boson, which is the only missing piece in the standard model of electroweak interactions, has been recently discovered
at the Large Hadron Collider (LHC) at the mass between $125~{\rm GeV}$ and $126.5~{\rm GeV}$~\cite{:2012gu,:2012gk}.  The newly discovered boson at LHC is found to behave much like in many ways the standard model Higgs particle, though there are a few anomalies in its decay modes such as the enhancement in the two photon channel or the suppression in the $b\,\bar b$ channel. 

Whether the anomalies hint new physics beyond the standard model or not is yet to
to be seen.  However, one strongly believes that there should be 
new physics beyond SM that  necessarily contains new particles to explain naturally the unknown component of matter in the Universe, dark matter~\cite{Jungman:1995df}.  The successful formation of large scale structures  and the temperature anisotropy in the cosmic microwave background require a sizable density perturbation of dark matter  in the much earlier epoch than the last scattering of photons. Dark matters produced in the early Universe survive until today and they show their existence in the several astrophysical phenomena
such as galactic rotation curves, gravitational lensing~\cite{Bertone:2004pz}. The common solution for the origin of  Higgs and dark matter may lie in a new physics beyond the standard model.

Models of technicolor (TC) theory~\cite{Weinberg:1975gm} assume a new strong dynamics of new particles, called techni-fermions,  at around $1~{\rm TeV}$ to break the electroweak symmetry dynamically. Higgs boson is then a bound state, constituted of techni-fermions, below the scale of 
techni-fermion condensation, induced by the new strong interaction.  TC therefore solves  the hierarchy problem naturally. 
Depending on the TC dynamics, Higgs boson could be a broad resonance around $1~{\rm TeV}$ just like $\sigma$ meson in QCD or 
a light and narrow bound state~\cite{Hong:2004td}, difficult to be distinguished from elementary Higgs at low energy. 
The weak gauge bosons get the masses through Higgs mechanism, when 
the TC becomes strong and techni-fermions condense, breaking  the electroweak gauge symmetry spontaneously.  
For the SM fermion masses, however,  one needs to introduce an additional new interaction, 
called extended TC (ETC) interaction~\cite{Dimopoulos:1979es}, mediated by massive particles at some scale $\Lambda_{\rm ETC}$, 
much higher than the electroweak scale, 
to transmit the techni-fermion condensation to the SM fermions, yielding their masses through 
the induced Yukawa couplings as in the SM~\cite{Farhi:1980xs}.

The original TC scenario based on a naive scale-up version of QCD was excluded:  
ETC scale is required to be at least around $300$ - $1000\,{\rm TeV}$~\cite{Farhi:1980xs} in order to suppress  
the induced flavor changing neutral currents between the SM fermions. Compared to the scale of the techni-fermion condensate, such ETC scale is, however,  
too high to generate enough mass to strange and other heavy quarks. 
This problem was solved, however, by modifying TC dynamics to be non-QCD like, generating a large anomalous dimension~\cite{Holdom:1981rm} for
the techni-fermion bilinear $\gamma_m \simeq 1$~\cite{Yamawaki:1985zg}, which enhances 
the techni-fermion condensate enough to account for SM fermion masses except top quark mass, 
for which one may introduce other mechanism~\cite{Miransky:1988xi,Miransky:1988gk,Matumoto:1989hf}. 
It was shown~\cite{Yamawaki:1985zg} that the large anomalous dimension $\gamma_m \simeq 1$ can indeed be realized   
when the TC gauge coupling $\alpha$ exhibits almost nonrunning behavior in the chirally broken phase 
$\alpha > \alpha_c$, where $\alpha_c$ is the critical coupling for the chiral symmetry breaking.  
The almost nonrunnning behavior actually implies the existence of 
a `quasi' infrared fixed point (IRFP), denoted as $\alpha_*$, known as 
the Caswell-Banks-Zaks IR fixed point (CBZ-IRFP)~\cite{Caswell:1974gg}, which is 
very close to but slightly larger than $\alpha_{c}$ 
so that $\alpha$ remains almost constant, $\alpha_{\rm c}\lesssim\alpha<\alpha_*$, 
exhibiting an approximate scale symmetry for wide range of scales. 
But, at scales lower than the dynamical techni-fermion mass ($E<m_F$), 
the techni-fermions decouple and the TC coupling runs quickly toward infinity and confines techni-gluons. (See Fig.~\ref{beta}.) 
TC with such a `quasi' IRFP is nowadays termed as walking TC (WTC)~\cite{Holdom:1981rm,Yamawaki:1985zg,Akiba:1985rr}. 
It is expected that the quasi IRFP emerges in cases with a large number of techni-fermions in the fundamental representation 
or a small number of techni-fermions in the higher dimensional representations~\cite{Hong:2004td,Sannino:2004qp}. 
Current lattice simulations support both possibilities~\cite{lattice}. 
WTC has been shown by several theoretical approaches~\cite{Appelquist:1991is,Hong:2006si} 
to be compatible with the electroweak precision data and thus serve as a viable framework for physics beyond the SM.

If WTC is a model for new physics at TeV, it is desirable to have a candidate for dark matter within the model.  
The lightest techni-baryon has been a popular candidate for dark matter~\cite{Nussinov:1985xr}, since 
the lightest techni-baryon is absolutely stable up to the nonperturbative anomalous decay just like ordinary QCD baryons. 
The techni-baryons could be either fermionic or bosonic, depending on the number of TCs $(N_{\rm TC})$. 
This scenario, however, needs an extra mechanism, some like the techni-baryon asymmetry, to account for its abundance in the present Universe.  

Recently another interesting candidate for dark matter is proposed~\cite{Choi:2011fy}:
If the electroweak symmetry breaking sector is highly conformal, 
techni-dilaton (TD), the (pseudo) Nambu-Goldstone boson, associated with spontaneously broken scale symmetry, is very light and weakly coupled to become a good candidate for dark matter. 
When the scale symmetry is spontaneously broken due to the techni-fermion condensation,  by Goldstone theorem 
WTC should have a dilaton as a Nambu-Goldstone boson associated with the scale symmetry. If WTC is extremely conformal, moreover, which can be achieved by adjusting the `new physics' such that the intrinsic ultraviolet 
scale of TC, $\Lambda_{\rm TC}\,(\gg \Lambda_{\rm ETC})$, lies very close to the IR fixed point or $\alpha(\Lambda_{\rm TC})\approx\alpha_*$ (See Fig.~\ref{beta}), 
the TD decay constant $F_{\rm TD}$ can be much bigger than the electroweak scale $v_{\rm EW}$, 
$\eta=v_{\rm EW}/F_{\rm TD}\ll1$  
so that the TD interacts extremely weakly with the strength suppressed by $1/F_{\rm TD}$ and its mass $M_{\rm TD}$ becomes much smaller 
than the techni-fermion mass scale $m_F$.  
This observation has been actually supported by a theoretical analysis on TD~\cite{Haba:2010hu} 
to suggest a critical scaling in the extremely walking limit, 
$M_{\rm TD}/m_F \to 0$ and $F_{\rm TD}/m_F \to \infty$ as $\alpha\, (\simeq \alpha_*)\to \alpha_c$~\footnote{
In Ref.~\cite{Haba:2010hu} the extremely walking limit has been quoted as a phenomenologically 
uninteresting limit in a sense that the TD gets decoupled from the SM particles, 
so cannot be seen at LHC. 
Though it may be irrelevant to the LHC physics, the extremely walking limit actually leads to 
an astrophysically and cosmologically interesting scenario, as was previously reported in Ref.~\cite{Choi:2011fy} and will also 
be seen more explicitly later in this paper. }. 
The TD in the extremely WTC therefore could be an interesting candidate for dark matter. 
In Ref.~\cite{Choi:2011fy}, indeed, the authors explored a possibility for a very light TD (what we call a light decoupled TD) 
to be a good candidate for dark matter and showed that 
the light decoupled TD indeed explains the observed relic abundance consistently with several cosmological and 
astrophysical constraints.

This paper will provide the detail of calculations done in Ref.~\cite{Choi:2011fy} 
and present more thorough analyses on the light decoupled TD as a dark matter candidate.  
This paper is organized as follows: 
In Sec.~\ref{review} we briefly review WTC as a model for beyond the SM and derive the TD couplings to the SM particles. 
We also discuss briefly  the properties of composite Higgs in TC to fit current LHC data. 
In Sec.~\ref{cosmo} we discuss the cosmological production of the light decoupled TD as a dark matter in the early Universe. 
The cosmological and astrophysical constraints on the light decoupled TD are discussed in Sec.~\ref{constraint} and 
its detection in the laboratory is discussed in Sec.~\ref{detection}. 
Summary of our paper is given in Sec.~\ref{summary}\,.

\section{Walking Technicolor and Techni-dilaton} 
\label{review}

In this section we shall discuss essential properties of TD 
as a pseudo Nambu-Goldstone boson of the spontaneous breaking of 
the approximate scale invariance. First, in Sec.~\ref{WTC} we recapitulate the salient features of WTC, 
needed for our discussion, viewed as the approximate scale invariance.  
We review the partially conserved dilatation current (PCDC) 
and derive the related formulas, relevant to later discussions, in Sec.~\ref{PCDC:sec}. 
We then find that the characteristic features, arising from scaling at the criticality 
in the WTC, allow an extremely light TD to be present in the Universe. 
The TD couplings with SM gauge bosons and fermions as well as techni-fermions 
are given in Secs.~\ref{TD:couplings:g} and~\ref{TD:couplings:y}. 
In Sec.~\ref{lifetime:sec} we estimate the lifetime for the light decoupled TD 
and give a cosmological bound on the TD mass and decay constant necessary for the light decoupled TD to be 
a dark matter. 
The generic difference between TD and a SM-like composite Higgs are explained and the constraints of composite Higgs due to recent discovery of a Higgs-like boson are
discussed in Sec.~\ref{compositeHiggs}.

\subsection{Walking technicolor} 
\label{WTC}

TC introduces a new strong dynamics at TeV energy scale to  break the electroweak symmetry dynamically. 
When the TC interaction becomes strong,  techni-fermions, new particles having the TC charges but transforming just like 
ordinary SM fermions under the SM interactions, form condensates, which then spontaneously break the electroweak symmetry.  
Higgs particle therefore arises in TC as a radial excitation of the condensate. 
\begin{figure}[t] 
\begin{center} 
\includegraphics[width=0.4\textwidth]{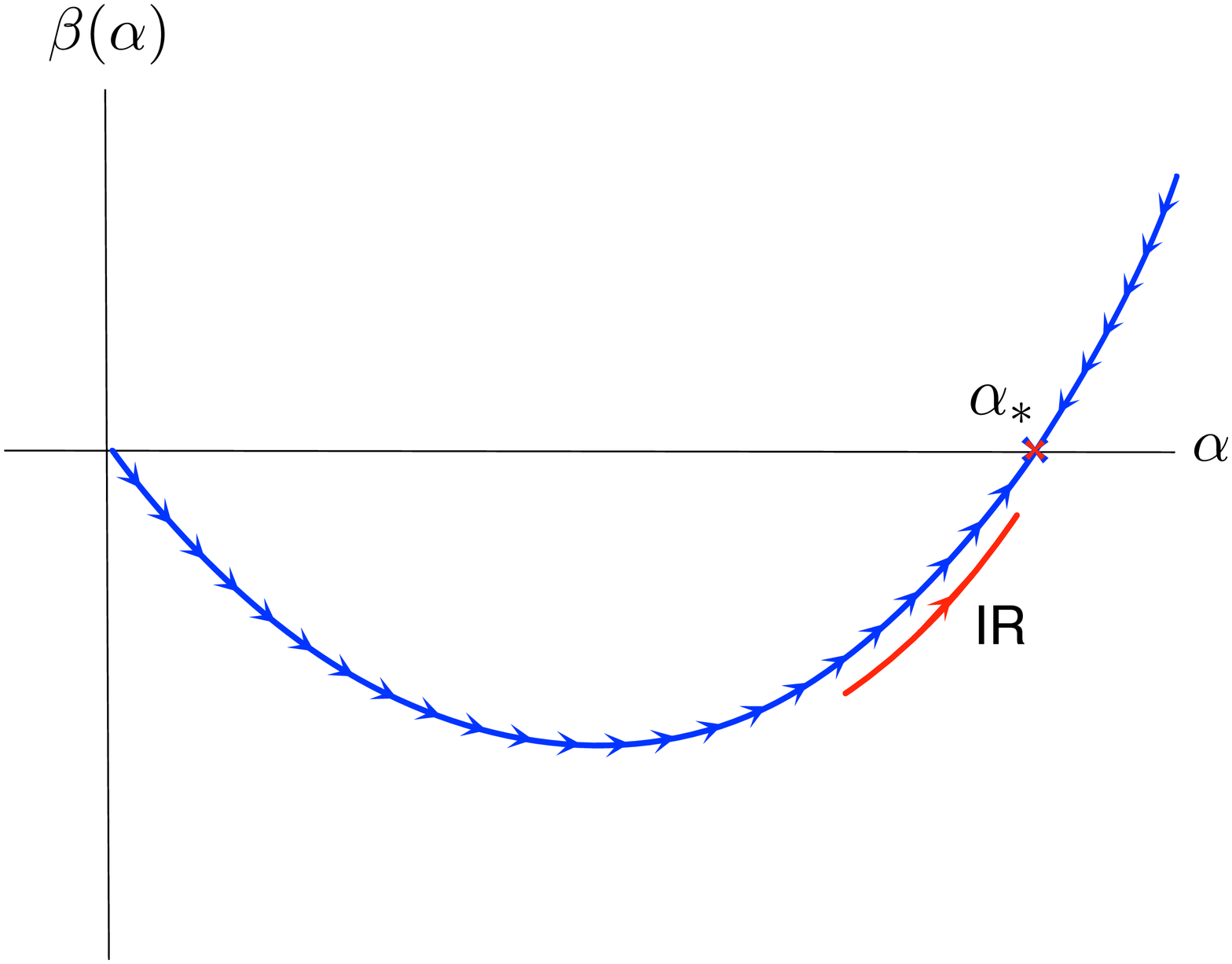}
\hskip 0.5in \includegraphics[width=0.45\textwidth]{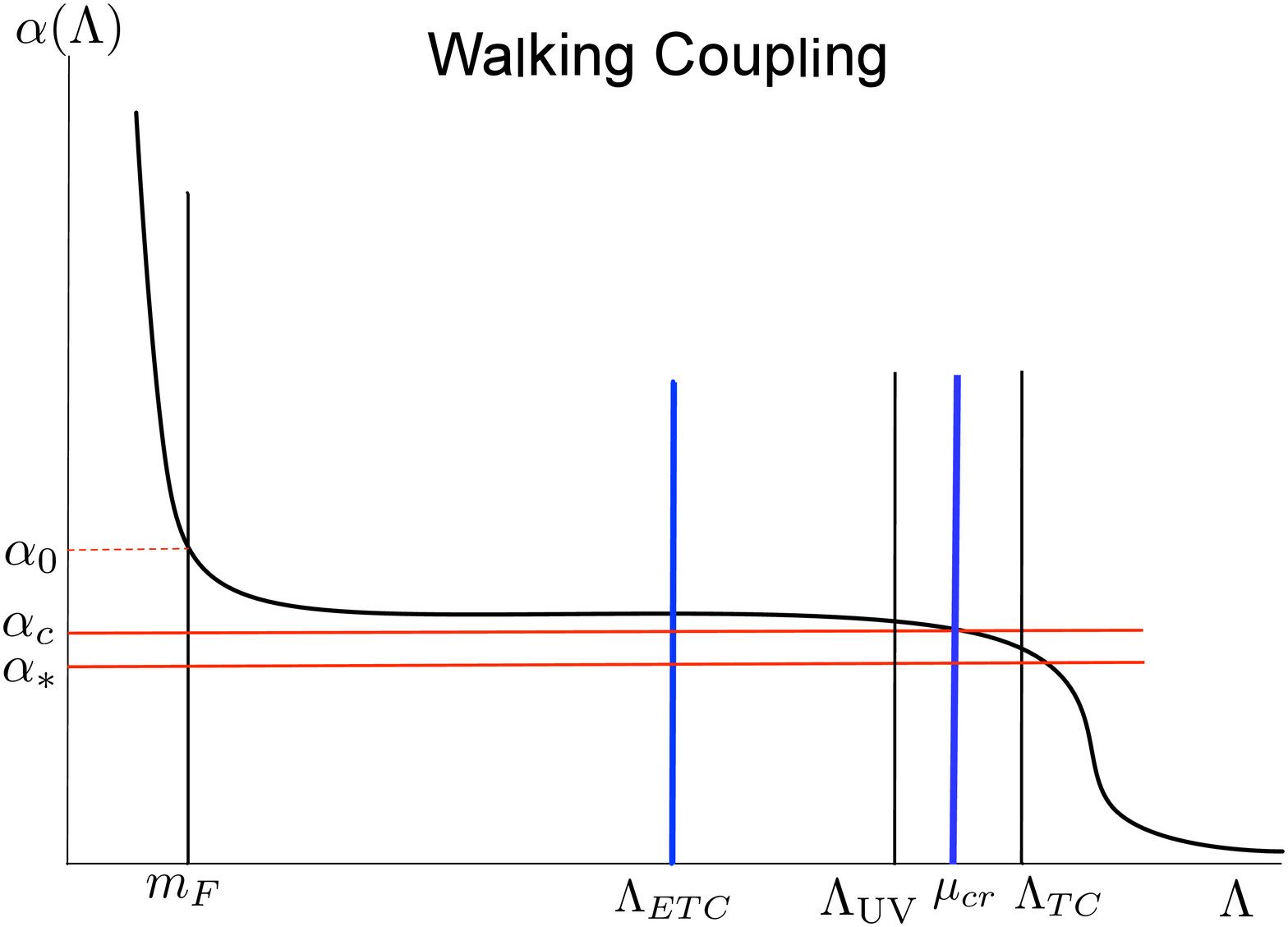}
\end{center}
\caption{ The beta function of TC coupling $\alpha$ with the CBZ-IRFP $\alpha_*$ (left panel) and the scale-dependence of TC 
coupling (right panel).} 
\label{beta}
\end{figure}
The CBZ-IRFP at which the beta function for the TC coupling vanishes, 
$\beta(\alpha_*)=0$, is present in TC 
with large number of techni-fermions in fundamental representations 
or a few number of those in high-dimensional representations, 
which provides one of the concrete dynamics to realize the walking. 
We further assume that the critical coupling, $\alpha_c$ for chiral symmetry breaking is very close 
to but slightly smaller than the CBZ-IRFP so that the TC coupling remains almost constant for a wide range of scale. 
(See Fig.~\ref{beta}.)

The chiral phase transition at $\alpha=\alpha_c$ of WTC is known as a (quantum) conformal phase transition~\cite{Miransky:1996pd} and exhibits Berezinsky-Kosterlitz-Thouless (BKT)~\cite{BKT} or Miransky scaling~\cite{Miransky:1984ef}:
\begin{equation}
m_F\approx\Lambda_{\rm UV}\,\exp\left(-\frac{\pi}{\sqrt{\alpha/\alpha_c-1}}\right)\,,
\label{miransky}
\end{equation}
where $m_F$ is the dynamical techni-fermion mass in the broken phase, $\alpha$ is the TC coupling measured at $\Lambda_{\rm UV}$,  
a ultraviolet (UV) scale of TC, chosen to be in the range of scales where $\alpha$ exhibits the walking behavior, 
as shown in Fig.~\ref{beta}. 
Having the critical coupling for chiral symmetry breaking very close to $\alpha_*$, 
the WTC generates the large mass hierarchy dynamically due to the quantum conformal phase transition at $\alpha_c$.

In the walking region $m_F\lesssim \Lambda
\lesssim\Lambda_{\rm UV}$ and $\alpha_0=\alpha(m_F) \gtrsim  \alpha >   \alpha_c$ the approximate scale invariance is present 
and hence the Bethe-Salpeter (BS) equation for a bound state of techni-fermion and anti-techni-fermion, $Q_{\rm TC}$, takes the following form:
\begin{equation}
\left[P^2+\partial^2+\frac{\alpha/\alpha_c}{r^2}\right]\chi_P(x)=0,
\end{equation}
where the amplitude of the bound state with momentum $P$ is defined at the origin of its center of mass coordinate as 
\begin{equation}
\chi_P(2x)=\left<0\right|T\, Q_{TC}\left(x\right)\bar Q_{TC}\left(-x\right)\left|P\right>
\end{equation}
and we used the ladder approximation since the vertex corrections are negligible in the walking region. 
The kernel for the BS equation in the ladder approximation  is given just by one techni-gluon exchange and is singular at short distances. Introducing a cutoff and analytically continuing to the Euclidean space, we regularize the kernel as, $a\to0$,
\begin{equation}
V(r)=\begin{cases} -\frac{\alpha/\alpha_c}{r^2} & \text{if $r\ge a$,}
\\
-\frac{\alpha/\alpha_c}{a^2} &\text{if $r\le a$.}
\end{cases}
\end{equation}
Requiring the bound state energy to be independent of the cutoff, $a$, we find a nonperturbative running coupling~\cite{Hong:1989zza}
\begin{equation}
\alpha(\Lambda)=\alpha_c+\alpha_c\frac{{\pi}^2}{\left[\ln\left(\frac{\Lambda}{\mu}\right)\right]^2}\,,
\end{equation}
where $\Lambda=a^{-1}$ is the UV cut-off and $\mu$ is a scale, generated by the dimensional transmutation due to the nonperturbative running.
The nonperturbative beta function is then
\begin{equation}
\beta_{\rm NP}(\alpha)=\Lambda\frac{\partial}{\partial \Lambda}\alpha(\Lambda)=-\frac{2\alpha_c}{\pi}\left(\frac{\alpha}{\alpha_c}-1\right)^{3/2}\,.
\label{np}
\end{equation}
 In the supercritical phase ($\alpha>\alpha_c$) the gap equation for the techni-fermions has a nontrivial and consistent solution, if the above nonperturbative beta function (\ref{np}) is employed~\cite{Miransky:1984ef,Leung:1985sn}. 
 Since the dynamical mass of techni-fermion should be renormalization-group (RG) invariant, we find 
\begin{equation}
m_F\simeq\Lambda({\alpha})\exp\left[-\int_{\alpha_0}^{\alpha}\frac{d\alpha^{\prime}}{\beta_{\rm NP}(\alpha^{\prime})}\right]\approx\Lambda_{\rm UV}\,e^{-\frac{\pi}{\sqrt{\alpha/\alpha_c-1}}}\,,
\end{equation}
where $\alpha$, the TC coupling at $\Lambda_{\rm UV}$, is much closer to $\alpha_c$ than $\alpha_0$, 
the coupling at $m_F$ (See Fig.~\ref{beta}). For the WTC, 
$\alpha$ is very close to $\alpha_c$ and the dynamical mass is indeed extremely small, 
compared to the UV scale of TC. 
Since  the anomalous dimension for the techni-fermion bilinear, $\gamma_m\simeq1+\sqrt{\alpha/\alpha_c-1}$, 
is very close to 1 in the extreme walking ($\alpha\to\alpha_c$),  
there emerges a marginal four-Fermi operator at low energies ($E\ll\Lambda_{\rm UV}$). 
As noted in \cite{Nonoyama:1989dq, Hong:1989zza}, 
the newly generated scale, $m_F$, in the WTC is associated to the dimensional transmutation of the 
dimensionless coupling of the marginal four-Fermi operator. 
(See also \cite{Kaplan:2009kr} for a recent discussion on this.)

\subsection{Techni-dilaton and its couplings}

In this subsection we make a brief review of properties of TD as the pseudo Nambu-Goldstone boson 
of the spontaneously broken scale symmetry and derive its couplings to the SM gauge bosons and fermions 
as well as techni-fermions.

\subsubsection{Partially conserved dilatation current}  

\label{PCDC:sec}

We begin by defining the decay constant of TD, $F_{\rm TD}$, as 
\begin{equation} 
  \langle 0 | D^\mu(x)  |{\rm TD}: p \rangle \equiv 
- i F_{\rm TD} p^\mu e^{-ipx} 
\,,  \label{FTD:def}
\end{equation}
or equivalently, 
\begin{equation} 
   \langle 0 | \theta^{\mu \nu}(x)  |{\rm TD}: p \rangle \equiv 
\frac{F_{\rm TD}}{3}  (p^\mu p^\nu - p^2 g^{\mu\nu}) e^{-ipx} 
\,,  \label{FTD:def2}
\end{equation}
where $D_\mu(x)$ is the dilatation current 
and $\theta^{\mu \nu}$ denotes the symmetric part of conserved energy-momentum tensor 
related to $D_\mu$ as $D_\mu=\theta_{\mu\nu} x^\nu$. 
Because the scale invariance is approximate, 
the TD gets its mass $M_{\rm TD}$: 
Acting the derivative on the both sides of Eq.(\ref{FTD:def}) or operating $g_{\mu\nu}$ in Eq.(\ref{FTD:def2}) 
one gets 
\begin{equation} 
   \langle 0 | \partial_\mu D^\mu(x)  |{\rm TD}: p \rangle 
= 
   \langle 0 | \theta_\mu^\mu(x)  |{\rm TD}: p \rangle 
=  
-  F_{\rm TD} M_{\rm TD}^2 e^{-ipx} 
\,, \label{PCDC:00}
\end{equation}
where $p^2=M_{\rm TD}^2$. 
Assuming the hypothesis of  partially conserved dilatation currents (PCDC), we obtain an approximate relation that the divergence of dilatation current is proportional to an interpolating TD field, 
\begin{equation} 
  \partial^\mu D_\mu (x)= \theta_\mu^\mu(x) 
= - F_{\rm TD} M_{\rm TD}^2 D(x) 
  \,, \label{PCDC:01}
\end{equation}
with the TD field $D(x)$ satisfying $\langle 0| D(x) | {\rm TD}: p \rangle = e^{-ipx}$.

We next consider the Ward-Takahashi identity 
regarding the dilatation current 
to combine it with the PCDC relation in Eq.(\ref{PCDC:00}) or Eq.(\ref{PCDC:01}). 
We start with the following matrix element: 
\begin{equation} 
  {\cal M}^\mu(q) \equiv \int d^4x \, e^{iqx} \langle 0| {\rm T} D^\mu(x) \theta_\nu^\nu(0)|0 \rangle
\,. \label{Mmu}
\end{equation} 
Multiplying both sides of Eq.(\ref{Mmu}) 
by $q_\mu$ and using $[i D^0(0,{\vec x}), \theta_\mu^\mu(0)]= \delta^{(3)}(\vec x) \delta_D \theta_\mu^\mu(0)$ 
with $\delta_D$ being an infinitesimal shift by the scale transformation,  
we obtain the Ward-Takahashi identity for the almost-conserved dilatation current or $M_{\rm TD}\approx0$:  
\begin{equation} 
q_\mu {\cal M}^\mu (q) 
\approx \langle 0| \delta_D \theta_\mu^\mu |0 \rangle 
= d_{\theta} \langle 0| \theta_\mu^\mu |0 \rangle 
\,, \label{RHS}
\end{equation}
where $d_{\theta} = 4$ is the scaling dimension of the energy-momentum tensor.  
On the other hand one can calculate  the product $q_\mu {\cal M}^\mu(q)$, assuming the TD pole dominance near $q^2=M_{\rm TD}^2$,
\begin{equation}
M^{\mu}(q)\approx \int d^4x\,e^{iqx}\langle 0|D^{\mu}(x)\left(\int_{p^2\approx M_{\rm TD}^2}|{\rm TD}\,:p\rangle\frac{i}{p^2-M_{\rm TD}^2}\langle {\rm TD}\,:p|\right)\,\theta_{\nu}^{\nu}(0)\rangle\,.
\end{equation}  
Taking the low-energy limit $q_\mu \to 0$ but $q^2\gg M_{\rm TD}^2\approx0$, we find 
\begin{equation} 
\lim_{q_\mu \to 0} 
q_\mu {\cal M}^\mu(q) = F_{\rm TD} \langle {\rm TD}:q=0 | \theta_\mu^\mu  | 0\rangle 
\,, 
\end{equation}
where we used  Eq.(\ref{FTD:def}). 
Comparing this with Eq.(\ref{RHS}) we have 
\begin{equation} 
 \langle {\rm TD}:q=0 | \theta_\mu^\mu | 0\rangle  
 = \frac{4}{F_{\rm TD}} \langle 0| \theta_\mu^\mu |0 \rangle 
 \,. 
\end{equation}
Using this and taking the corresponding amplitude for Eq.(\ref{PCDC:01}),   
we thus rewrite the PCDC relation Eq.(\ref{PCDC:01}) as 
\begin{equation} 
F_{\rm TD}^2 M_{\rm TD}^2 = - 4 \langle0| \theta_\mu^\mu |0\rangle = - 16\, {\cal E}_{\rm vac} 
\,, \label{PCDC:0}
\end{equation} 
where ${\cal E}_{\rm vac} = \langle 0 | \theta_0^0 |0 \rangle$ 
denotes the vacuum energy density.

\begin{figure}[t] 
\begin{center} 
\includegraphics[scale=0.4]{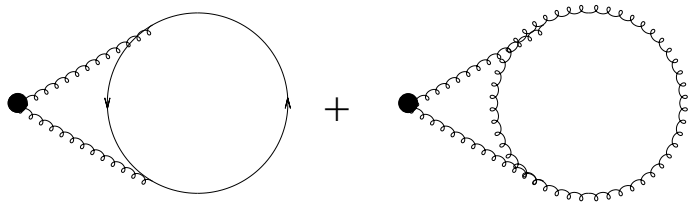}
\end{center}
\caption{ The blob denotes the insertion of the trace of the TC energy momentum tensor. The wiggly line denotes the full techni-gluon propagator and the solid line denotes the full techni-fermion propagator. } 
\label{np-vac}
\end{figure}

The vacuum energy density contains all the contributions both from TC  particles and SM particles.
Near the quasi IR fixed point, however, the perturbative contributions to the vacuum energy coming from both TC and SM particles are negligible because the beta functions almost vanish.   
On the other hand, because of the scale anomaly due to non-perturbative beta function, Eq.~(\ref{np}), in the TC sector 
the vacuum energy gets contributions from the techni-gluon condensation (See Fig.~\ref{np-vac}): 
\begin{equation} 
\langle 0 | \partial_\mu D^\mu | 0 \rangle  
= 4 {\cal E}_{\rm vac} 
= \langle 0| \frac{\beta(g_{\rm TC})}{2 g_{\rm TC} } (G^{\rm TC}_{\mu\nu})^2     |0\rangle 
\,. 
\end{equation} 
The vacuum energy density ${\cal E}_{\rm vac}$ has been evaluated in Refs.~\cite{Hashimoto:2010nw,Miransky:1989qc}. 
We here adopt an extremely walking case which is well simulated by nonrunning (standing) 
limit of $g_{\rm TC}$. 
The vacuum energy ${\cal E}_{\rm vac}$ 
is then calculated through the Cornwall-Jackiw-Tomboulis effective potential and 
is found to be dominated by the techni-fermion loop~\cite{Miransky:1989qc}:
\begin{equation} 
    {\cal E}_{\rm vac}
   = -  \frac{N_{\rm TC} N_{\rm TF}}{\pi^4} m_F^4 
   \,, \label{CJT}
\end{equation}
where $N_{\rm TC}$ and $N_{\rm TF}$ respectively stand for the number of TC and that of techni-fermions. 
 Combining Eq.(\ref{CJT}) with Eq.(\ref{PCDC:0}) 
we finally arrive at a concise PCDC formula, 
\begin{equation} 
  F_{\rm TD}^2 M_{\rm TD}^2 = \frac{16 N_{\rm TC} N_{\rm TF}}{\pi^4} m_F^4 
\, , \label{PCDC}
\end{equation} 
to which we will hereafter refer as the PCDC relation.

Since the WTC yields $\gamma_m \simeq 1$ for the techni-fermion bilinear operator $\bar{F}F$, 
the induced four-Fermi operator $(\bar{F}F)^2$ having ${\rm dim}(\bar{F}F)^2 \simeq 4$ becomes marginal as well as the TC gauge coupling $\alpha$ 
in the sense of renormalization group analysis. 
The form of scale anomaly, hence the PCDC relation (\ref{PCDC}), should then 
be modified by the presence of the four-Fermi interaction. 
As was briefly noted in Sec.~\ref{WTC}, 
such four-Fermi effects have been intensively studied through the analysis on the 
planar QED with nonrunning gauge coupling and four-Fermion interactions added 
(what is called gauged Nambu-Jona-Lasinio (NJL) model)~\cite{Bardeen:1985sm,Leung:1985sn,Nonoyama:1989dq,Shuto:1989te,Leung:1989hw,Bardeen:1991sv,Carena:1992cg}. 
Particularly in Refs.~\cite{Nonoyama:1989dq,Shuto:1989te}, 
the vacuum energy density ${\cal E}_{\rm vac}$ was explicitly computed in the gauged NJL model 
with the nonrunning gauge coupling, so that the result essentially remains the same as in Eq.(\ref{PCDC}), 
${\cal E}_{\rm vac} \sim N_{\rm TC}N_{\rm TF} m_F^4$. 
Even for a perturbatively running case, it leads 
to essentially the same result on ${\cal E}_{\rm vac}$ as that in Eq.(\ref{PCDC})~\cite{Hashimoto:2010nw} 
within a 5\% uncertainty. 
This reflects the fact that the mass and coupling of TD are tied to the nonperturbative scale anomaly which 
has nothing to do with how the theory is perturbatively (fully or almost) scale invariant, as long as the 
dynamical fermion mass is generated in accord with {\`a} la Miransky scaling (\ref{miransky}).

\subsubsection{The Yukawa interactions} 
\label{TD:couplings:y}

Following the standard procedure as done in Ref.~\cite{Bando:1986bg}, we derive the Yukawa couplings 
between fermions and TD 
arising from the techni-fermion condensation and ETC contributions. 
We start with the Ward-Takahashi identity about the dilatation current coupled to 
techni-fermion bilinear $\bar{F}F$: 
\begin{eqnarray} 
 q^\mu {\cal M}_\mu (q) &=& - (3-\gamma_m) \langle 0| \bar{F} F  |0\rangle 
\,, \nonumber \\ 
 {\cal M}_\mu(q) &=& \int d^4 x\, e^{iqx} \langle 0| {\rm T} D_\mu(x) \bar{F}(0) F(0)|0 \rangle 
 \,, \label{WT:FF}
\end{eqnarray}
where $(3-\gamma_m)$ denotes the scaling dimension of $\bar{F}F$ including the anomalous dimension $\gamma_m$. 
 Assuming the TD pole dominance in the left hand side of Eq.(\ref{WT:FF}), we evaluate it in 
the low-energy limit $q_\mu \to 0$ to get 
\begin{equation} 
\lim_{q_\mu \to 0} q^\mu {\cal M}_\mu (q)  = F_{\rm TD}
\langle {\rm TD}:p|  \bar{F} F |0 \rangle 
\,.  
\end{equation} 
 Comparing this with the right hand side of Eq.(\ref{WT:FF}) we have 
 \begin{equation} 
\langle {\rm TD}:p=0|  \bar{F} F |0 \rangle 
= - \frac{3-\gamma_m}{F_{\rm TD}} \langle 0| \bar{F} F  |0\rangle 
\,.  
\end{equation} 
This implies that  
\begin{equation} 
  \bar{F}F \approx \langle 0| \bar{F}F  |0\rangle 
- (3-\gamma_m)\langle 0| \bar{F}F  |0\rangle \frac{D}{F_{\rm TD}} 
\,.  \label{WT:yukawa}
\end{equation} 
We may consider four-Fermi interaction terms: 
\begin{equation} 
  {\cal L}_{\rm 4-fermi} 
  = G_1 \bar{F} F \bar{F} F + G_2 \bar{F}F \bar{f} f 
  \,, 
\end{equation}
which would be generated by exchange of a ``communicator" between TC and SM sectors, like ETC gauge boson. 
These terms yield fermion masses via the techni-fermion condensation: 
\begin{equation} 
  m_{F,f} = - G_{1,2} \langle 0| \bar{F} F |0 \rangle 
  \,. 
\end{equation}
 Combining these with Eq.(\ref{WT:yukawa}) we find the Yukawa interaction terms 
between TD and techni-($F$), SM ($f$) fermions, 
\begin{equation} 
 {\cal L}^{\rm Yukawa} 
 = -  \frac{1}{\sqrt{2}} D  \sum_{f, F} 
\left( g_{DFF} \bar{F} F + g_{Dff} \bar{f} f \right) 
 \,, \label{Yukawa:int}
\end{equation} 
with the Yukawa couplings $g_{DFF}$ and $g_{Dff}$  
\begin{eqnarray} 
 \frac{g_{D FF}}{\sqrt{2}} &=& 
(3-\gamma_m) \frac{m_F}{F_{\rm TD}}  
 \nonumber \\ 
  \frac{g_{D ff}}{\sqrt{2}} &=& 
(3-\gamma_m) \frac{m_f}{F_{\rm TD}} 
\,. \label{yukawa}
\end{eqnarray}

\subsubsection{The couplings to gauge bosons}  
\label{TD:couplings:g}

The TD couplings to the SM gauge bosons are generated through techni-fermion loops.  
In a low energy region with $p < m_F$, 
the gauge interactions between the TD field $D$ and $SU(3)_c$, $SU(2)_W$ and $U(1)_Y$ gauge boson fields 
$(G_\mu, W_\mu, B_\mu)$ then take the form: 
\begin{eqnarray}
{\cal L}_{\rm gauge} 
&=& 
\frac{2(3-\gamma_m)}{F_{\rm TD}} v_{\rm EW}^2 D\, {\rm tr}[g_W W_\mu - g_Y B_\mu]^2
- \frac{\beta(g_s)}{2g_s} \frac{(3-\gamma_m)}{F_{\rm TD}} D (G_{\mu\nu})^2 \nonumber \\ 
&&  
- \frac{\beta(g_W)}{2g_W } \frac{(3-\gamma_m)}{F_{\rm TD}} D (W_{\mu\nu})^2  
- \frac{\beta(g_Y)}{2g_Y } \frac{(3-\gamma_m)}{F_{\rm TD}} D (B_{\mu\nu})^2   
\,, \label{gaugeint:anomalous}
\end{eqnarray}
where $v_{\rm EW} \simeq 246$ GeV and the beta functions are 
defined as $\beta(g_i) = \frac{g_i^3}{(4\pi)^2} b_i$ for $i={\rm TC}, s, W,Y$ 
with the beta function coefficients $b_i$. 
As will be discussed later, 
the $D-G-G$ term in Eq.(\ref{gaugeint:anomalous}) become relevant when the TD thermal 
production processes are evaluated.

\subsection{ The lifetime for a long-lived techni-dilaton} 
\label{lifetime:sec}

 \begin{figure}[t] 

\begin{center} 
\includegraphics[scale=0.8]{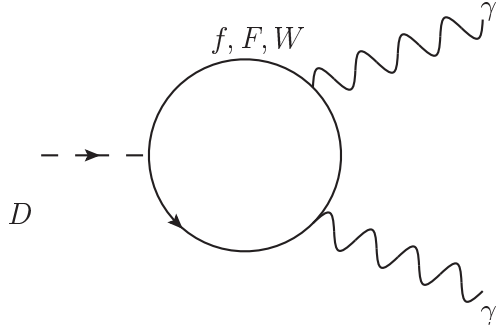}
\end{center}
\caption{ The diagram for the TD decay into two photons through techni-($F$), the standard model $(f)$ fermion, and $W$ boson loops. 
 For $W$ boson loop contribution, the graph displayed here involves  
another type of loop constructed from the $W-W-\gamma-\gamma$ four-point vertex.  
} 
\label{TD-2gamma-graph}
\end{figure}

 In this subsection we shall discuss the TD decay modes and the lifetime 
in the case of a light decoupled TD in an extremely walking scenario.

The light decoupled TD is supposed to have the mass $M_{\rm TD} \ll m_F, m_f$, 
so that the decay rate is governed by two-photon decay mode  
$D \to \gamma\gamma$.  
The decaying process can involve both gauge and Yukawa interactions, 
Eqs.(\ref{Yukawa:int}) and (\ref{gaugeint:anomalous}), as depicted in Fig.~\ref{TD-2gamma-graph}. 
 Taking the limit $M_{\rm TD} \ll m_F, m_f$ we calculate the decay rate to get   
\begin{equation} 
 \Gamma(D \to \gamma\gamma)
\simeq 
 \frac{\alpha_{\rm EM}^2}{36 \pi^3} \frac{M_{\rm TD}^3}{F_{\rm TD}^2} \big|{\cal C}-\frac{5}{2}\big|^2
 \,, \label{decay-width2}
\end{equation} 
where $\alpha_{\rm EM}=e^2/(4\pi)$ and 
\begin{equation} 
{\cal C}= N_{\rm TC} \sum_F N_c^{(F)} Q_F^2 
\,, 
\end{equation} 
in which $N_c^{(F)}=3(1)$ for techni-quarks (leptons) and $Q_F$ the electric charge of $F$-techni-fermion. 
Using \eq{PCDC} 
we thus estimate the lifetime of TD, $\tau_{\rm TD}$, to get  
\begin{eqnarray} 
 \tau_{\rm TD} 
&\simeq&  \Gamma^{-1}(D \to \gamma\gamma) 
 \nonumber \\ 
&\simeq& 
10^{17} \, {\rm sec} \,  
 \left( \frac{10 \, {\rm keV}}{M_{\rm TD}} \right)^5 \left( \frac{m_F}{10^3 \, {\rm GeV}} \right)^4 
\,, 
 \label{lifetime}
\end{eqnarray} 
 for the one-family model~\cite{Farhi:1980xs} with $N_{\rm TC}=2$ which gives ${\cal C}=16/3$.

 One might think that 
$D \to \bar{\nu}\nu$ decay rate can be included in the estimate of $\tau_{\rm TD}$ 
since Eq.(\ref{lifetime}) implies $M_{\rm TD} > 2 m_\nu$, which, 
however, turns out to be negligible due to the large suppression factor  
$(m_{\nu}/(\alpha_{\rm EM}M_{\rm TD}))^2 \sim 10^{-10}$ compared to the decay rate of Eq.(\ref{decay-width2}).

For TD to be a dark matter, the lifetime in Eq.(\ref{lifetime}) has to be longer than the age of the Universe, $\sim 10^{17}$ sec, 
which places an upper bound for the TD mass, 
\begin{equation} 
M_{\rm TD} \lesssim 10 \, {\rm keV}
\,, 
\end{equation}
for the one family model with $N_{\rm TC}=2$ and $m_F=10^3$ GeV. 
This constraint also gives a lower bound on the TD decay constant through Eq.(\ref{PCDC}), 
\begin{equation} 
F_{\rm TD}\gtrsim10^{11}\gev 
\,, \label{FTD:11}
\end{equation} 
which indeed implies the decoupled TD. 
The extremely large $F_{\rm TD}$ as in Eq.(\ref{FTD:11}) suggests to us that 
$F_{\rm TD}$ arises as the cutoff scale of WTC, $\Lambda_{\rm TC}$, namely, 
\begin{equation} 
  F_{\rm TD} \simeq \Lambda_{\rm TC} \gtrsim 10^{11} {\rm GeV} 
  \,, 
\end{equation}  
which is, however, not necessarily identified with ETC scale that would be much higher or may not be present: 
 If $\Lambda_{\rm ETC} > 10^6$ GeV, the ETC exchange would yield too small mass for the SM fermions (e.g. strange quark)  
even in the case of WTC. 
The issue on the fermion mass generation is beyond the scope of this paper to be explored elsewhere.

\subsection{Techni-dilaton and composite Higgs}
\label{compositeHiggs}

As both TD and Higgs boson have the same quantum numbers, spin-0, positive parity, and charge neutral, 
they do mix with each other and have similar physical properties at low energy. 
It is therefore not easy to disentangle them at colliders. 
However, they are two different objects, created by totally different operators. 
TD is a Nambu-Goldstone boson, created out of the vacuum by dilatation currents, 
while Higgs in TC is a composite field, created by the techni-fermion bilinear. 
In this subsection we present the comparison of TD with the composite Higgs in TC 
and also discuss their mixing.   In the decoupling limit, $F_{\rm TD}\gg m_F$, that we are interested in, the mixing between the techni-dilaton and the composite Higgs is extremely small and thus negligible. 

Since a Higgs-like boson of mass around $125~{\rm GeV}$ has been discovered at LHC, it is natural to assume that the composite Higgs in our model is indeed the new particle discovered at LHC. We then briefly discuss the constraints on the couplings of composite Higgs to the standard model particles.

\subsubsection{Techni-dilaton-Higgs coupling}
We define a composite field $\Phi$, made of techni-fermion bilinear field $\bar{F}{F}$ to discuss the composite Higgs in TC,
\begin{equation}
\Phi(x)=\lim_{y\to x}\frac{\left|x-y\right|^{\gamma_m}}{v^2} F(y)\bar F(x)\,,
\end{equation}
where $v$ and $\gamma_m$ are the vacuum expectation value and the anomalous dimension of the techni-fermion bilinear, defined earlier, respectively. The composite Higgs couples to SM fields through interactions with techni-fermions, 
so the TD couples to the composite Higgs field as well:  
In general, dilaton couples to operators of any fields as long as their scaling dimension differs from 4. 
(Being a scalar graviton, dilaton couples universally.) 
To see the TD coupling to the composite Higgs explicitly, 
we consider the following matrix element of a product of dilatation current $D_\mu$ and Higgs field $\Phi$: 
\begin{equation}
{\cal H}_{\mu}(q)=\int{d^4x}\,e^{iqx}\langle 0|{\rm T}D_{\mu}(x)\,\Phi^{\dagger}\Phi(0)|0\rangle\,.
\end{equation}
Just as before we multiply the external momentum, $q^\mu$, and assume the dilaton pole dominance at low energy ($q^2\to0$) to 
get the Ward-Takahashi identity which provides the TD coupling to the composite Higgs: 
\begin{equation}
2\langle 0|\Phi^{\dagger}\Phi(0)|0\rangle=F_{\rm TD}\langle {\rm TD}:q=0|\Phi^{\dagger}\Phi(0)|0\rangle\,.
\label{higgs-WT}
\end{equation}
In a low-energy effective Lagrangian, Eq.(\ref{higgs-WT}) can be viewed as 
the following dilaton-Higgs coupling in the Higgs mass term: 
\begin{equation}
{\cal L}\ni -\frac12m_H^2\left(1+\frac{2D}{F_{\rm TD}}\right)\Phi^{\dagger}\Phi\,,
\end{equation}
which is nothing but the leading two terms in the scale-invariant Higgs mass term in the effective Lagrangian, 
\begin{equation}
{\cal L}_{\rm  mass}=-\frac12m_H^2e^{2D/F_{TD}}\Phi^{\dagger}\Phi\,.
\label{higgs-dilaton}
\end{equation}

\subsubsection{Mixing}
\begin{figure}[t] 
\begin{center} 
\includegraphics[scale=0.5]{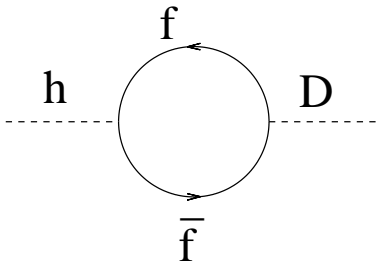}
\end{center}
\caption{ The mixing between the physical Higgs and techni-dilaton through a fermion loop. } 
\label{mixing}
\end{figure}

Once the composite Higgs develops its condensate, we may write the Higgs field in the unitary gauge as 
\begin{equation}
\Phi(x)=\begin{pmatrix}
	     0\\
	     v_{\rm EW}+h(x)
	     \end{pmatrix}\,.
\end{equation}
In addition to the tree level coupling (\ref{higgs-dilaton}),  the physical Higgs $h(x)$ and TD mix with each other
through fermion loops as depicted in  Fig.~\ref{mixing}. 
The mass matrix of Higgs and techni-dilaton is generated at one-loop level as 
\begin{equation}
(h, D)\begin{pmatrix}
m_h^2& -\Delta\\
-\Delta & m_{TD}^2
\end{pmatrix}
\begin{pmatrix}
h\\
D
\end{pmatrix}\,,
\end{equation}
where $\Delta=\frac{3-\gamma_m}{8\pi^2v_{\rm EW}F_{\rm TD}}\sum_f m_f^4$. 
Writing the mass eigenstates $h^{\prime},D^{\prime}$  as 
\begin{equation}
h^{\prime}=h\cos\phi+D\sin\phi,\quad D^{\prime}=D\cos\phi-h\sin\phi\,, 
\end{equation}
we find the size of the mixing angle $\phi$ is set by $v_{\rm EW}/F_{TD}$, which turns out to be quite 
small, $v_{\rm EW}/F_{\rm TD} \sim 10^{-9}$, for our extremely walking scenario. 
The mixing between TD and the composite Higgs is therefore negligible in our discussions.

\subsubsection{Potential energy}
The potential energy for the composite Higgs can be calculated from, for instance, Cornwall-Jackiw-Tomboulis formalism and one can expand it in powers of Higgs fields. All terms, allowed by the symmetry, should be present:
\begin{equation}
V(\Phi)=\frac12m_H^2\Phi^{\dagger}\Phi+\frac{\lambda}{4}\left(\Phi^{\dagger}\Phi\right)^2+\cdots,
\end{equation}
where the couplings, $m_H^2$, $\lambda$, $\cdots$ are calculable in principle.
On the other hand the potential for the TD is dictated by the scale anomaly~\cite{Migdal:1982jp}:
\begin{equation}
V(\chi)=\left|{\cal E}_{\rm vac}\right|\chi^4\left(4\ln\chi-1\right)\,, 
\label{TD:p}
\end{equation}
where $\chi=e^{D/F_{\rm TD}}$\,. 
The forms of their potentials are thus quite different so could be testable at some future collider experiments 
having high luminosity. 

\subsubsection{Composite Higgs of $125~{\rm GeV}$}
Since the composite Higgs in TC is similar to $\sigma$ particle in QCD, one might assume its mass is around $m_F\sim1~{\rm TeV}$, the infrared (IR) scale of WTC,   and has a broad width. However, WTC has a very different dynamics and thus its spectrum would be quite different from that of QCD, following the Miransky scaling~\cite{Miransky:1996pd}. Indeed, it was shown in the large $N_{\rm TC}$ limit the composite Higgs can be as light as  $150~{\rm GeV}$ and has a narrow width~\cite{Hong:2004td}. 

\begin{figure}[t] 
\begin{center} 
\includegraphics[scale=0.37]{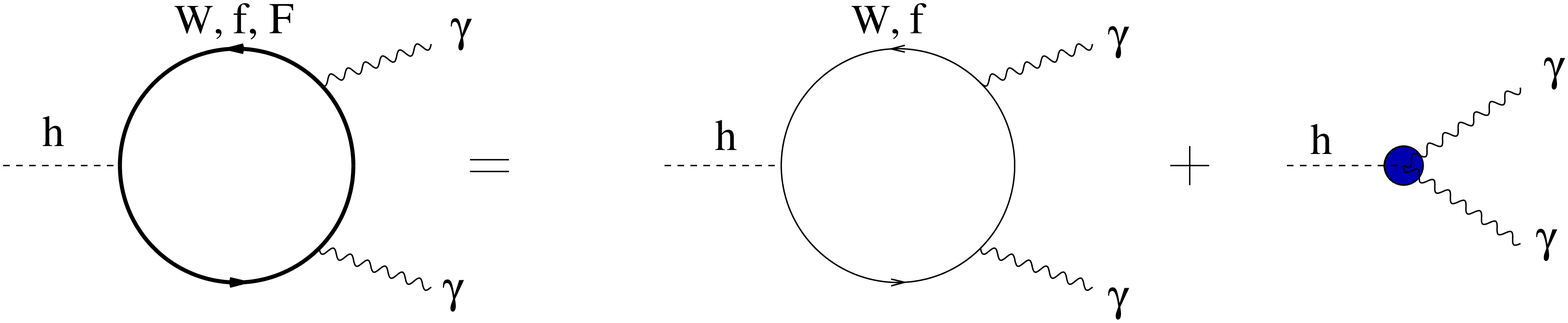}
\end{center}
\caption{ The leading coupling of Higgs to photons, where $W$ bosons, both SM fermions ($f$) and techni-fermions ($F$) contribute. The blob denotes the new contribution due to TC.} 
\label{composite}
\end{figure}

The couplings of composite Higgs to the SM fermions and the longitudinal components of the $W$ and $Z$ bosons are same as those of elementary SM Higgs, since TC is constructed such that. However, its couplings to photons and gluons will be model-dependent. The leading effective interactions of (composite) Higgs to photons and gluons are given as 
\begin{eqnarray}
{\cal L}_{\rm eff}\ni C_{\gamma} \frac{h}{v_{\rm ew}}F_{\mu\nu}F^{\mu\nu}+C_{g} \frac{h}{v_{\rm ew}}G^a_{\mu\nu}G^{a\mu\nu}+{\rm h.o.}\,.
\end{eqnarray}
The diagram for the effective coupling of (composite) Higgs to photons is shown in Fig~{\ref{composite}}. (The diagram for the effective coupling to gluons will be similar.) 
The effective coupling has two pieces, one due to  SM particles and another due to TC:
\begin{equation}
C_{\gamma}=C_{\gamma}^{\rm SM}+\delta C_{\gamma}\,.
\end{equation}
(The coupling due to SM particles is same as that of SM Higgs.) Similarly the effective couplings to gluons will be 
\begin{equation}
C_{g}=C_{g}^{\rm SM}+\delta C_{g}\,.
\end{equation}
Since the cross section for the Higgs production via gluon fusion and decay into two photons is proportional to 
\begin{equation}
C^2_{\gamma}\cdot C_{g}^2\,,
\end{equation}
we need $C_g\cdot C_{\gamma}\approx 1.3\, C_g^{\rm SM}\cdot C^{\rm SM}_{\gamma}$ upto a sign to 
fit the current excess in the two-photon channel~\cite{:2012gu,:2012gk}. 

\begin{figure}[t] 
\begin{center} 
\includegraphics[scale=0.5]{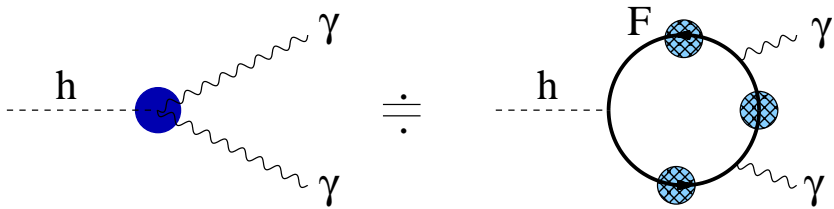}
\end{center}
\caption{ The TC contributions to the  Higgs coupling to photons in the ladder approximation. The blob in the techni-fermion (F) loop denotes the self-energy of techni-fermions due to TC interactions.} 
\label{ladder}
\end{figure}

In SM the dominant contributions to the Higgs coupling to photons come from the $W$ boson loop. But, in TC we do not know how big is $\delta C_{\gamma}$, not to mention its sign. Since the $\beta$ function of technicolor is small for WTC models, one may use the ladder approximation in calculating the TC contributions to the Higgs coupling, neglecting the vertex corrections, which is tantamount to having massive techni-fermions in the loop for low external momenta. 
(See Fig.~\ref{ladder}.)
Therefore, the TC contribution to the Higgs coupling to photons is similar to that of massive top quark. For the one-family TC model, then, the TC contribution is $2\,N_{\rm TC}$ times that of top quark. We find that  $\left(C_{\gamma}^{\rm SM}+\delta C_{\gamma}\right)^2=1.76\,(C_{\gamma}^{\rm SM})^2$ for $N_{\rm TC}=8$. 
Therefore it is not unreasonable to get 30\% enhancement in the Higgs coupling to photons for models of WTC to account for 
the enhancement in the two-photon channel of Higgs decay. Currently, however, other processes like Higgs to $WW$ and $ZZ$ are consistent with SM Higgs. We therefore need the production rate of Higgs via gluon fusion to be similar to that of SM. Namely $\delta C_{g}\approx 0$~\footnote{In SM the top loop dominates in the Higgs-gluon coupling. But in TC we do have not only colored techni-quark but also colored techni-pions whose contributions might cancel each other. }, while $C_{\gamma}^2\approx 1.7 \, (C^{\rm SM}_{\gamma})^2$ in models of WTC to fit the current data.

Another interesting feature of current data on the newly discovered particle at LHC is slight suppression in the $b\,\bar b$ and $\tau^+\,\tau^-$ decay channels. At the tree level the (composite) Higgs couplings to SM fermions are same as that of SM Higgs, 
but the loop correction due to TC interactions could be different. It will be interesting to estimate the loop correction precisely, since such effect might be within the reach of $8~{\rm TeV}$ running at LHC.  
In summary, WTC models are roughly consistent with current data, but it is quite necessary to calculate precisely the effective couplings of Higgs to test WTC models\footnote{One attempt in this direction will be to use gauge/gravity duality or to rely on the lattice calculation.}, which will be done somewhere else.

\section{Cosmological production} 
\label{cosmo}

In this section we shall address cosmological production of TD in the early Universe; the thermal (Sec.\ref{TP}) 
and non-thermal productions (Sec.~\ref{NTP}). 
It turns out that the thermal production tends to be suppressed due to 
the decoupling features tied with the large TD decay constant, 
so that the non-thermal production becomes dominant to be large enough for the light decoupled TD 
to explain the present dark matter density.

\subsection{Thermal production} 
\label{TP}

The TD will be generated at the same time or right after 
the techni-fermion condensate takes place 
at the temperature $T=\mu_{\rm cr}$ where $\alpha = \alpha_{\rm cr}$ and 
$\mu_{\rm cr}$ satisfies $m_F ={\cal O}(10^3\, {\rm GeV}) < \mu_{\rm cr}<  
\Lambda_{\rm TC} (\simeq F_{\rm TD}) = {\cal O}(10^{11}\, {\rm GeV})$. 
As noted above, in addition, due to the Miransky scaling Eq.(\ref{miransky}), 
the dynamical mass $m_F$ should be much 
smaller than the other two scales, so that $m_F/\mu_{\rm cr} \ll 1$ and 
$m_F/\Lambda_{\rm TC} \ll 1$, i.e., $m_F \ll \mu_{\rm cr} < \Lambda_{\rm TC}$. 
This hierarchical structure has actually been confirmed~\cite{Hashimoto:2010nw} 
by an explicit calculation in the walking TC, which predicts 
$m_F/\Lambda_{\rm TC} \simeq 10^{-9}$ and $\mu_{\rm cr}/\Lambda_{\rm TC} \simeq 10^{-3}$ 
for the case with an {\it extremely large scale hierarchy}. 
We thus see that, due to the large decay constant $F_{\rm TD}\gtrsim 10^{11}$ GeV, 
the TD decouples (with the decoupling temperature $T_d \sim 10^{10}$ GeV)  
from the thermal equilibrium as soon as it is produced at $T=\mu_{\rm cr}< F_{\rm TD}$. 

Though the TD decouples  from the thermal equilibrium right after its generation, 
it could be produced through scatterings of particles which are 
in the thermal equilibrium with $T \lesssim \mu_{\rm cr}$, similar to the thermal 
production of E-WIMPs (extremely weakly interacting massive particles)~\cite{Choi:2005vq,Bailly:2009pe,Choi:2011yf}. 
Those thermal particles are assumed to include techni-hadrons and techni-fermions with masses of ${\cal O}(10^3 \, {\rm GeV})$ 
as well as the SM particles. 
The most dominant contribution is expected to come from processes 
including QCD interactions with the relatively large QCD coupling $\alpha_s \sim 0.1$. 
In the following we shall discuss such QCD processes based on an effective Lagrangian 
induced from techni-fermion loops, perturbatively expanding the amplitudes 
in powers of $\alpha_s$~\footnote{
Actually, the dynamics around $T=\mu_{\rm cr}$ is quite nonperturbative and hence all the 
TD couplings should be written as non-local form factors. 
It might therefore be unreliable if we work on an effective Lagrangian described only by 
local operators. However, our conclusion on the thermal production drown in this section 
will not essentially change even if such non-local contributions could be all taken into account, 
since the smallness of the thermal production is closely tied with the extremely large TD decay 
constant. }. 
It turns out that the thermal production is too small to accommodate the realistic dark matter relic density 
essentially due to the extremely small TD couplings suppressed by $1/F_{\rm TD}$.

In some class of WTC models we may have QCD color-charged techni-pions arising as 
pseudo Nambu-Goldstone bosons with mass of ${\cal O}(1\, {\rm TeV})$,  
which are bound states of anti-techni-quark and -lepton, $\bar{Q}L$ belonging to ``triplet", {\bf 3}, of QCD, 
or anti-techni-quark and techni-quark $\bar{Q}Q$ belonging to ``octet",{\bf 8}, of QCD.  
These spectra can generically be classified by basis of $SU(2)$-weak isospins and (techni-) baryon number charge $U(1)_V$, labeled in total 
by $I=0,1,2,3$. We shall denote them as $P_3^I$ ($\bar{P}_3^I$) for triplet-(anti-)pions and $P_8^I$ for octet-pions.  
The colored-techni-pion terms coupled with TD are generated from techni-fermion triangle loops 
in the same fashion as depicted in Fig.~\ref{TD-2gamma-graph}. 
Those loop-induced vertices are collected into 
the following effective Lagrangian terms: 
\begin{eqnarray} 
  {\cal L}_{P_3}^{\rm inv} &=& \sum_n^{N_{P_3}} \sum_I \frac{2(3-\gamma_m)}{F_{\rm TD}} D (D_\mu \bar{P}_{3n}^I D^\mu P_{3n}^I)  
\,, \nonumber  \\
  {\cal L}_{P_8}^{\rm inv} &=& \sum_n^{N_{P_8}} \sum_I \frac{2(3-\gamma_m)}{F_{\rm TD}} D ({\rm Tr}[D_\mu P_{8n}^I D^\mu P_{8n}^I ] )
  \,, \label{lag:inv}
\end{eqnarray}
where $N_{P_3}$ and $N_{P_8}$ respectively denote the number of color-triplet and -octet techni-pions with isospin $I$, 
and  
\begin{eqnarray} 
  D_\mu P_3^I &=& \partial_\mu P_3^I - ig_s G_\mu P_3^I 
\,, \\ 
     D_\mu P_8^I &=& \partial_\mu P_8^I - ig_s [G_\mu, P_8^I] 
\,,  
\end{eqnarray}
with $P_3^{I} = P_3^{Ii}$ ($i=1,2,3$) and 
$P_8^I = P_8^{I\alpha} (\frac{\lambda^\alpha}{2})_{ij}$ ($\alpha=1,\cdots 8$).

\begin{figure}

\begin{center} 
\includegraphics[scale=0.7]{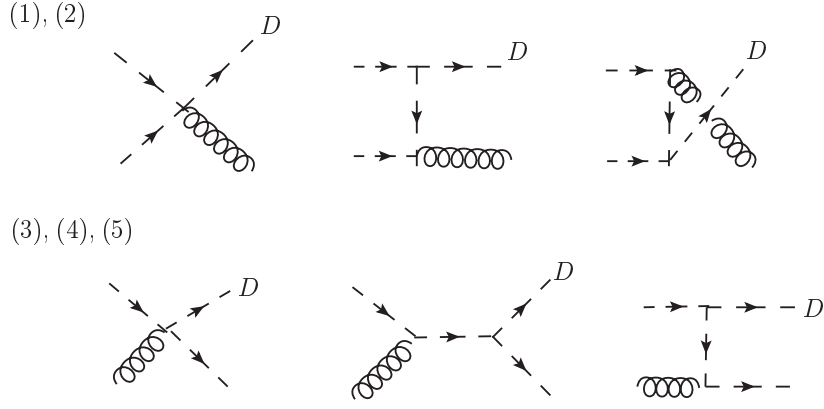}
\end{center}
\caption{The Feynman graphs corresponding to the TD production processes listed in Eq.(\ref{process1}). 
The wavy and dotted lines denote gluons and colored-techni-pions, receptively.  }
\label{leading-graphs-TD}
\end{figure}

  From the interaction terms in Eq.(\ref{lag:inv}) we find that 
the leading contribution comes from single TD production processes 
as illustrated in Fig.~\ref{leading-graphs-TD}: 
\begin{eqnarray} 
\begin{array}{llcl} 
(1) & P_{3n}^{Ii} + \bar{P}_{3n}^{Ij} &\to& D + g^\alpha \\ 
(2) & P_{8n}^{I\alpha} + P_{8n}^{I\beta} &\to& D +g^\gamma \\
(3) & P_{3n}^{Ii} + g^\alpha &\to& D + P_{3n}^{Ij}  \\ 
(4) & \bar{P}_{3n}^{Ii} + g^\alpha &\to& D + \bar{P}_{3n}^{Ij}  \\ 
(5) & P_{8n}^{I\alpha} + g^\beta &\to& D + P_{8n}^{I\gamma} 
\end{array} 
\,. \label{process1}
\end{eqnarray} 
We calculate these amplitudes in the massless limit for all the particles, 
which is a good approximation for $m_F \le T \le \mu_{\rm cr}$ that we are interested in.   
 We then find  that at the leading order (LO) in $\alpha_s/F_{\rm TD}^2$  
the amplitude vanishes for each process: 
\begin{equation} 
i {\cal M} (a+ b \to D + c) \Bigg|_{\rm LO} = 0 
\, \qquad 
{\rm with} \quad a,b,c \, \in \, (1)-(5) 
 \,. \label{LO:amp}
\end{equation}

\begin{figure}

\begin{center} 
\includegraphics[scale=0.5]{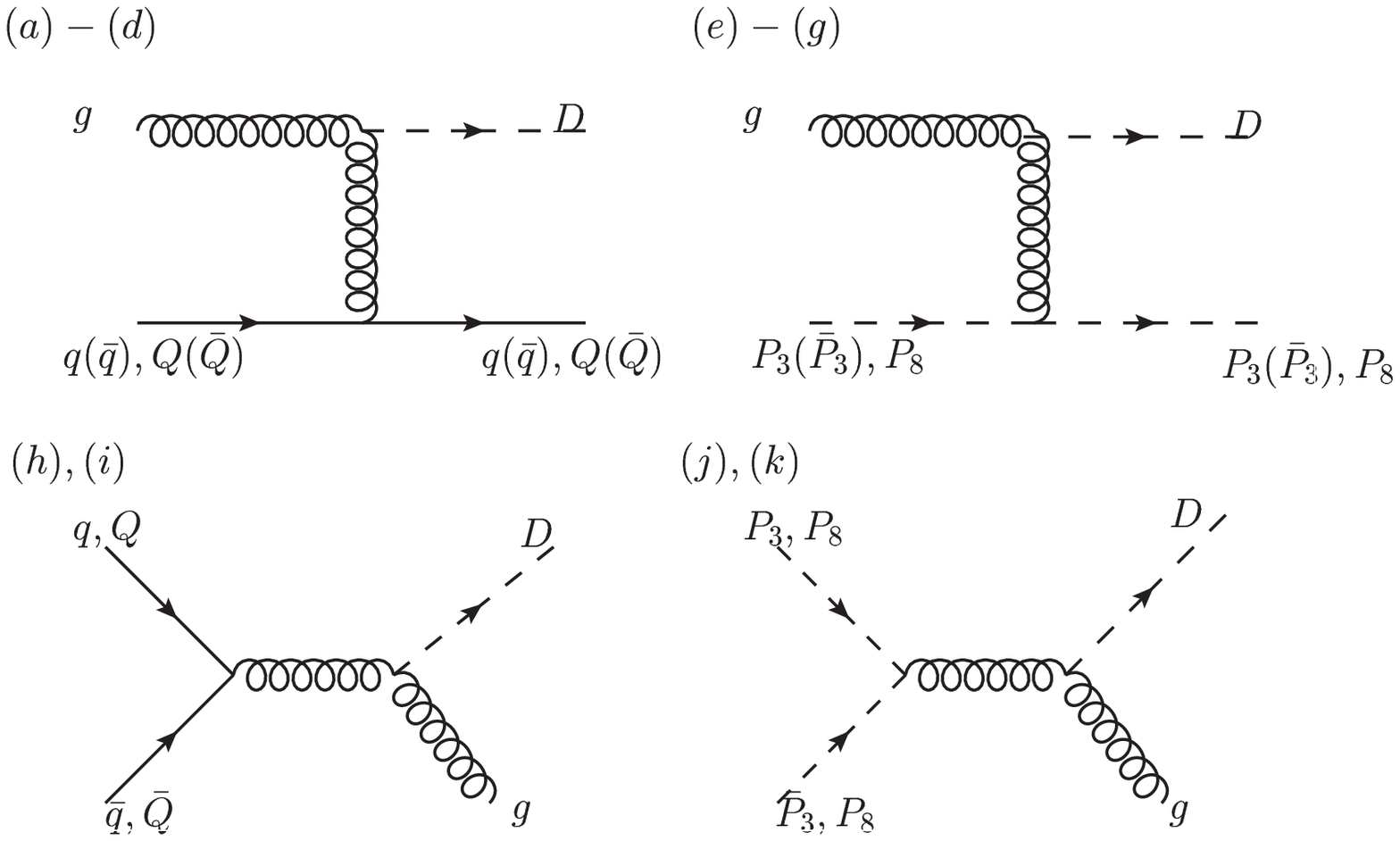}
\end{center}
\begin{center}
\includegraphics[scale=0.5]{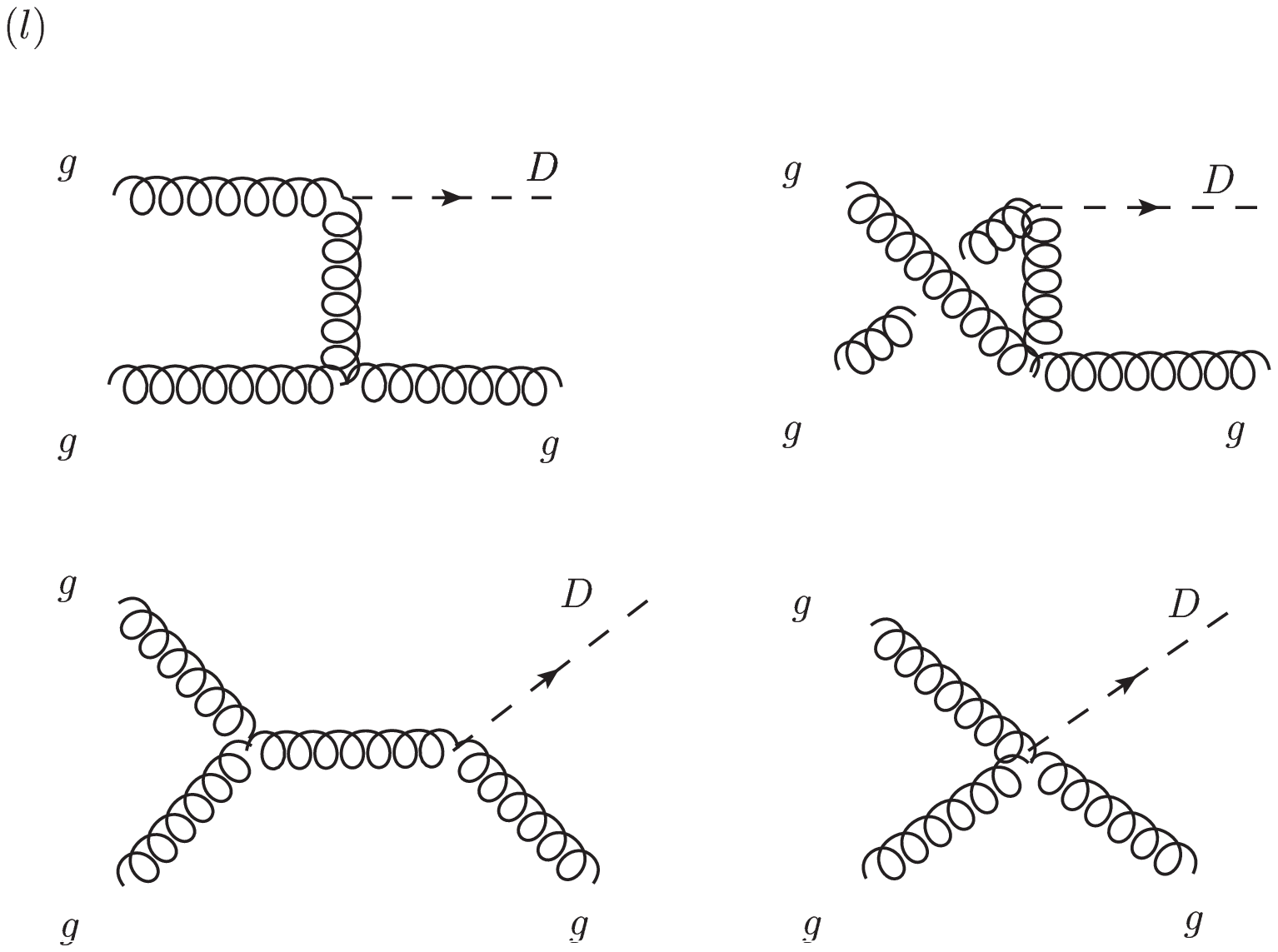}
\end{center}
\caption{The Feynman graphs corresponding to the TD production processes listed in Eq.(\ref{process2}). }
\label{td-production}
\end{figure}

We shall next discuss the next to leading order contributions to the TD production. 
It turns out that those contributions arise from the $D-G-G$ vertex in Eq.(\ref{gaugeint:anomalous}): 
\begin{equation}
{\cal L}_{DGG} = - \frac{\beta(g_s)(3-\gamma_m)}{2g_s F_{TD}} D (G_{\mu\nu}^\alpha G^{\alpha \mu\nu})  
\,. \label{L:DGG}
\end{equation}
 From this term we construct 
the amplitudes yielding nonzero TD production having single TD in the final state 
along with the single suppression factor $(1/F_{\rm TD})$. 
The relevant Feynman graphs are depicted in Fig.~\ref{td-production} and the corresponding production processes are 
as follows: 
\begin{eqnarray} 
\begin{array}{llcl} 
(a) \hspace{15pt} 
& q + g^\alpha & \to & D + q \\ 
(b) & \bar{q} + g^\alpha &\to & D + \bar{q} \\ 
(c) & Q + g^\alpha & \to& D + Q \\ 
(d) & \bar{Q} + g^\alpha &\to& D +\bar{Q} \\ 
(e) & P_{3n}^{Ii} + g^\alpha &\to& D + P_{3n}^{Ij}  \\ 
(f) & \bar{P}_{3n}^{Ii} + g^\alpha &\to& D + \bar{P}_{3n}^{Ij}  \\ 
(g) & P_{8n}^{I\alpha} + g^\beta &\to& D + P_{8n}^{I\gamma} \\ 
(h) & q + \bar{q} &\to& D + g^\alpha \\ 
(i) & Q + \bar{Q} &\to& D + g^\alpha \\ 
(j) & P_{3n}^{Ii} + \bar{P}_{3n}^{Ij} &\to& D + g^\alpha \\ 
(k) & P_{8n}^{I\alpha} + P_{8n}^{I\beta} &\to& D +g^\gamma \\
(l) & g^\alpha + g^\beta &\to& D + g^\gamma 
\end{array} 
\,. \label{process2}
\end{eqnarray}
The squared amplitudes with all the quantum numbers in the initial and final states summed up 
are calculated to be  
\begin{eqnarray} 
\sum_{\alpha, q} \frac{1}{4} \sum_{\rm spins} |i {\cal M}_{(a),(b)}|^2 
  &=& \frac{(3-\gamma_m)^2\beta^2}{2 F_{\rm TD}^2}  N_{q} \left(\frac{N_c^2-1}{2} \right) \left[ 
 - \frac{s(s+t)}{t-m_g^2} - 2 t 
\right] 
\,, \\ 
\sum_{\alpha, Q} \frac{1}{4} \sum_{\rm spins} |i {\cal M}_{(c),(d)}|^2 
  &=& \frac{(3-\gamma_m)^2\beta^2}{2 F_{\rm TD}^2} N_Q N_{\rm TC} \left(\frac{N_c^2-1}{2} \right) \left[ 
 - \frac{s(s+t)}{t-m_g^2} - 2 t 
\right] 
\,, \\ 
\sum_{i,j, \alpha} \sum_{I,n} \frac{1}{2} \sum_{\rm spins} |i {\cal M}_{(e),(f)}|^2 
  &=& \frac{(3-\gamma_m)^2\beta^2}{4 F_{\rm TD}^2} 4 N_{P_3} \left(\frac{N_c^2-1}{2} \right)  \left[ 
 3t - \frac{4s(s+t)}{t-m_g^2} 
\right] 
\,, \\ 
\sum_{\alpha, \beta, \gamma, I,n} \frac{1}{2} \sum_{\rm spins} |i {\cal M}_{(g)}|^2 
  &=& \frac{(3-\gamma_m)^2\beta^2}{4 F_{\rm TD}^2} 4 N_{P_8}  N_c   \left( N_c^2-1 \right)  \left[ 
 3t - \frac{4s(s+t)}{t-m_g^2} 
\right] 
\,, \\ 
\sum_{\alpha, q} \frac{1}{4} \sum_{\rm spins} |i {\cal M}_{(h)}|^2 
  &=& \frac{(3-\gamma_m)^2\beta^2}{2 F_{\rm TD}^2} N_{q} \left(\frac{N_c^2-1}{2} \right) \left[ 
 \frac{t(t+s)}{s} + 2 s 
\right] 
\,, \\ 
 \sum_{\alpha, Q} \frac{1}{4} \sum_{\rm spins} |i {\cal M}_{(i)}|^2 
  &=& \frac{(3-\gamma_m)^2\beta^2}{2 F_{\rm TD}^2}  N_Q N_{\rm TC}  \left(\frac{N_c^2-1}{2} \right) \left[ 
 \frac{t(t+s)}{s} + 2 s 
\right] 
\,, \\ 
\sum_{i,j, \alpha, I,n} \sum_{\rm spins} |i {\cal M}_{(j)}|^2 
  &=& \frac{(3-\gamma_m)^2\beta^2}{2 F_{\rm TD}^2} 4 N_{P_3} \left(\frac{N_c^2-1}{2} \right)  \left[ 
 3s - \frac{4t(t+s)}{s} 
\right] 
\,, \\ 
\sum_{\alpha,\beta,\gamma, I,n} \sum_{\rm spins} |i {\cal M}_{(k)}|^2 
  &=& \frac{(3-\gamma_m)^2\beta^2}{2 F_{\rm TD}^2} 4 N_{P_8} N_c \left( N_c^2-1 \right)  \left[ 
 3s - \frac{4t(t+s)}{s} 
\right] 
\,, \\
  \sum_{\alpha, \beta, \gamma} \frac{1}{4} \sum_{\rm spins} |i {\cal M}_{(l)}|^2 
&=&  \frac{5 (3-\gamma_m)^2 \beta^2}{8 F_{\rm TD}^2} N_c (N_c^2-1) \left[ 
\frac{- t(s+t) (s^2 + st + t^2)^2}{s(m_g^2 - t)^2 (m_g^2 + s+t)^2} \right] 
 \,, 
\end{eqnarray}
where $N_q$ and $N_Q$ respectively denote the number of the standard model quarks and that of techni-quarks belonging to 
the fundamental representation of TC group. 
In calculating the squared amplitudes 
we took all the masses to be zero and 
supplied an infrared cutoff for the $t$-channel processes with the thermal gluon mass in plasma $m_g \sim g_s T$. 
Working at center-of-mass frame with the angle $\theta$ by which $t=-s/2 (1+\cos\theta)$, 
we evaluate the total cross section $\sigma = \int_{-1}^1 d \cos\theta \frac{|i {\cal M}|^2}{32\pi s}$ 
for each process to get 
\begin{eqnarray} 
\sigma_{(a), (b)}(s) 
&=& \frac{(3-\gamma_m)^2\beta^2}{4\pi F_{TD}^2} N_{q} \left( \frac{N_c^2-1}{2} \right) \left[ \frac{1}{8} \log \frac{s}{m_g^2}  \right] 
\,, \nonumber \\ 
\sigma_{(c), (d)}(s) 
&=& \frac{(3-\gamma_m)^2\beta^2}{4\pi F_{TD}^2} N_{Q} N_{\rm TC} \left( \frac{N_c^2-1}{2} \right) \left[ \frac{1}{8} \log \frac{s}{m_g^2}  \right] 
\,, \nonumber \\ 
 \sigma_{(e),(f)}(s) 
&=& \frac{(3-\gamma_m)^2\beta^2}{4\pi F_{TD}^2} N_{P_3} \left( \frac{N_c^2-1}{2} \right) \left[ \log \frac{s}{m_g^2} - \frac{11}{8}  \right] 
\,, \nonumber \\ 
 \sigma_{(g)}(s) 
&=& \frac{(3-\gamma_m)^2\beta^2}{4\pi F_{TD}^2} N_{P_8} N_c \left( N_c^2-1 \right) \left[ \log \frac{s}{m_g^2} - \frac{11}{8}  \right] 
\,, \nonumber \\ 
\sigma_{(h)}(s) 
&=& \frac{(3-\gamma_m)^2\beta^2}{4\pi F_{TD}^2} N_{q} \left( \frac{N_c^2-1}{2} \right) \left[ \frac{5}{24}  \right] 
\,, \nonumber \\ 
\sigma_{(i)}(s) 
&=& \frac{(3-\gamma_m)^2\beta^2}{4\pi F_{TD}^2} N_{Q} N_{\rm TC} \left( \frac{N_c^2-1}{2} \right) \left[ \frac{5}{24}  \right] 
\,, \nonumber \\
\sigma_{(j)}(s) 
&=& \frac{(3-\gamma_m)^2\beta^2}{4\pi F_{TD}^2} N_{P_3} \left( \frac{N_c^2-1}{2} \right) \left[ \frac{13}{6}  \right] 
\,, \nonumber \\ 
\sigma_{(k)}(s) 
&=& \frac{(3-\gamma_m)^2\beta^2}{4\pi F_{TD}^2} N_{P_8} N_c \left( N_c^2-1 \right) \left[ \frac{13}{6}  \right] 
\,, \\ 
  \sigma_{(l)}(s) 
  &=& \frac{(3-\gamma_m)^2\beta^2}{4\pi F_{\rm TD}^2} N_c (N_c^2-1) 
  \left[  \frac{5}{16} \log \frac{s}{m_g^2} - \frac{115}{192}  \right] 
  \,. \label{sigmas}
\end{eqnarray}

\subsubsection{The estimate of the thermally produced relic density of TD}

Now that we have obtained the explicit expressions of cross sections relevant to 
the thermal production of TD, we are  ready to estimate 
the relic density at present through the Boltzmann equation.

The Boltzmann equation for the TD number density $n_{\rm TD}$ 
is given by 
\begin{equation} 
  \frac{d n_{\rm TD}}{dt} + 3 H n_{\rm TD} = \sum_{i,j} \langle \sigma (i+j\to D + \cdots ) v \rangle n_i n_j   
  \,,\label{BEQ}
\end{equation}
where $H$ is the Hubble parameter $H(T) =  (\frac{\pi^2}{30} g_* T^4/3M_P^2)^{1/2}$ with 
the effective degrees of freedom $g_*(T)$ and we neglected inverse processes 
since the TD number density is much smaller than the photon number density in the thermal equilibrium. 
The thermal average $\langle \sigma (i+j\to D + \cdots ) v \rangle n_i n_j $ is expressed as~\cite{Choi:1999xm} 
\begin{equation} 
  \langle \sigma_n (i+j\to D + \cdots ) v \rangle n_i n_j 
  =  \zeta^2(3) \cdot \eta_i \eta_j  \cdot \frac{g_ig_j}{16\pi^4} T^6 \int_0^\infty dx x^4 K_1(x) \sigma(x^2) 
  \,, 
\end{equation} 
where $\zeta(3)= 1.202...$ 
and $K_1(x)$ denotes modified Bessel function, $\sigma(x^2) = \sigma(s/T^2)$ and $g_i$ is the internal (spin) degree of freedom 
of particle $i$; $\eta_i$ is a number density factor associated with the initial state particles 
assigned as $\eta_i=1(3/4)$ for bosons (fermions).

 It is convenient to introduce the TD density yield $Y_{\rm TD}\equiv \frac{n_{\rm TD}}{s}$ where $s$ denotes entropy density. 
We thus rewrite the Boltzmann equation (\ref{BEQ}) in terms of $Y_{\rm TD}$ using $H=\frac{1}{2t} \propto T^2$ and 
$s \propto T^3$ and then solve it with the boundary condition $Y_{\rm TD}(T=\mu_{\rm cr})=0$ to reach the formula for 
the TD yield at present $T=T_0$,   
\begin{eqnarray} 
  Y_{\rm TD}(T_0) 
  &=& \int_{T_0}^{\mu_{\rm cr}} dT \frac{\sum_{i,j} \langle \sigma (i+j\to D + \cdots ) v \rangle n_i n_j }{s(T) H(T) T}
\, \nonumber \\ 
 &=& 
 \frac{135\sqrt{10} M_P}{2 \pi^{3}} \int_{T_0}^{\mu_{\rm cr}} dT \frac{\sum_{i,j} \langle \sigma  (i+j\to D + \cdots ) v  \rangle n_i n_j}
{g_{*}^{3/2} (T)T^6}
 \,, 
\end{eqnarray}
where in reaching the last line we used 
$s=g_{s*}(T) \frac{2 \pi^2}{45} T^3$ with $g_*(T) = g_{s*}(T)$ assumed. 
Note that the TD production is highly suppressed below the techni-fermion mass scale $m_F$ because of 
the decoupling of techni-fermions and -pions, such that we may take 
$Y_{\rm TD}(T_0) \simeq Y_{\rm TD}(T=m_F)$.

\begin{table} 
\begin{tabular}{c|c|c|c|c|c} 
\hline
\hspace{10pt} $n$ \hspace{10pt}& \hspace{10pt}$\eta_i \cdot \eta_j$ \hspace{10pt}& \hspace{10pt}$g_i \cdot g_j$ \hspace{10pt} 
& \hspace{10pt} ${\cal N}$ \hspace{10pt}& \hspace{10pt} $A_n$ \hspace{10pt} & \hspace{10pt} $B_n$ \hspace{10pt}\\  
\hline\hline 
$(a)$, $(b)$ & $\frac{3}{4}$ & 
4  &  $N_q \left( \frac{N_c^2-1}{2} \right)$ & $\frac{1}{8}$ & $0$ \\  
\hline 
$(c)$, $(d)$ & $\frac{3}{4}$ & 
4  &  $N_Q N_{\rm TC}\left( \frac{N_c^2-1}{2} \right)$ & $\frac{1}{8}$ & $0$ \\  
\hline 
$(e)$, $(f)$ & 1 & 
2  &  $N_{P_3} \left( \frac{N_c^2-1}{2} \right)$ & $1$ & $-\frac{11}{8}$ \\  
\hline 
$(g)$ & 1 & 
2  &  $N_{P_8} N_c \left( N_c^2-1 \right)$ & $1$ & $-\frac{11}{8}$ \\  
\hline 
$(h)$ & $\frac{9}{16}$ & 
4  &  $N_{q} \left( \frac{N_c^2-1}{2} \right)$ & $0$ & $\frac{5}{24}$ \\  
\hline 
$(i)$ & $\frac{9}{16}$ & 
4  &  $N_{Q} N_{\rm TC}\left( \frac{N_c^2-1}{2} \right)$ & $0$ & $\frac{5}{24}$ \\  
\hline 
$(j)$ & $1$ & 
1 &  $N_{P_3} \left( \frac{N_c^2-1}{2} \right)$ & $0$ & $\frac{13}{6}$ \\  
\hline 
$(k)$ & $1$ & 
1 &   $N_{P_8} N_c \left( N_c^2-1\right)$ & $0$ & $\frac{13}{6}$ \\  
\hline 
$(l)$ & $1$ & 
4 &   $ N_c \left( \frac{N_c^2-1}{2} \right)$ & $\frac{5}{16}$ & $-\frac{115}{192}$ \\  
\hline\hline  
\end{tabular}
\caption{The list of numerical factors relevant to Eq.(\ref{Yn}). For the one-family model 
quoted in the text $N_q=6$, $N_c=3$, $N_Q=2$ and $N_{P_3}=N_{P_8}=1$.}  
\label{list}
\end{table}

To make the explicit estimate of $Y_{\rm TD}(T_0)$, 
we shall consider the one-family model with the number of TC $N_{\rm TC}$, techni-fermions $N_{\rm TF}=8 (N_Q=2)$. 
 For $m_F <T< \mu_{\rm cr}$ the QCD beta function $\beta$ in this model is explicitly given as 
\begin{equation} 
  \beta \Bigg|_{\rm one-family\, model} = - \frac{g_s^3}{(4\pi)^2} \left[ \frac{11}{3} - \frac{4}{3} N_{\rm TC} \right]
  \,. 
\end{equation} 
Taking into account the forms of the total cross sections in Eq.(\ref{sigmas}) 
we calculate $Y_{\rm TD}(T_0)$ in the one-family model to get  
\begin{equation} 
  Y_{\rm TD}(T_0) \Bigg|_{\rm one-family\, model} 
  \simeq 
  \frac{135 \sqrt{10} M_P}{512 \pi^9}  
\frac{(3-\gamma_m)^2 \left(
\frac{11}{3} - \frac{4}{3} N_{\rm TC} 
\right)^2}{F_{\rm TD}^2}  
\left( \int_{m_F}^{\mu_{\rm cr}} dT \frac{\alpha_s^3(T)}{g_{*}^{3/2}(T) } 
  \sum_{n=(a)-(l)} Y_{\rm TD}^{(n)} (T) \right) 
\,,  \label{Y:0}
\end{equation}
where 
\begin{equation} 
  Y_{\rm TD}^{(n)}(T) 
  = \zeta^2(3) \cdot \eta_i \eta_j \cdot g_i g_j \cdot {\cal N} \cdot \left[ A_n (2 I_1 - I_2 \log(4\pi \alpha_s(T))) + B_n I_2   \right]   
\,, \label{Yn}
\end{equation}
with 
\begin{eqnarray} 
  I_1 &=& \int_0^\infty dx x^4 K_1(x) \log x = 4(5-4 \gamma + \log 16) \simeq 21.85 
  \,, \\ 
  I_2 &=& \int_0^\infty dx x^4 K_1(x) = 16 
  \,, 
\end{eqnarray}
and other factors being listed in Table~\ref{list} for each process. 
Since $\mu_{\rm cr} \gg m_F$ and $\alpha_s$ runs only logarithmically during $m_F<T<\mu_{\rm cr}$, 
the dependence of $\mu_{\rm cr}$ on $Y_{\rm TD}(T_0)$ becomes almost linear.  
We thus numerically have 
\begin{eqnarray} 
  Y_{\rm TD}(T_0) \Bigg|_{\rm one-family\, model} 
&\simeq & 
\left( \frac{\mu_{\rm cr}}{10^8 {\rm GeV}} \right) 
\left(\frac{200}{g_*(\mu_{\rm cr})}  \right)^{3/2} 
\left( \frac{10^{11} {\rm GeV}}{F_{\rm TD}} \right)^2 
\nonumber \\ 
&& 
\times 
\Bigg\{  
\begin{array}{cc} 
2.2 \times10^{-3}  & \qquad {\rm for} \qquad N_{\rm TC}=2 \\  
9.4 \times 10^{-4}  & \qquad {\rm for} \qquad  N_{\rm TC}=3 
\end{array}
\,, \label{Y:0:onefamily}
\end{eqnarray} 
where we used $\alpha_s(m_F) \simeq \alpha_s (m_Z) \simeq 0.1$ 
and chose $\mu_{\rm cr}=10^{8}$ GeV inspired by 
the result of Ref.~\cite{Hashimoto:2010nw} 
where $\mu_{\rm cr}/\Lambda_{\rm TC} \simeq 10^{-3}$.

In Eq.(\ref{Y:0:onefamily}) we took $g_*(\mu_{\rm cr})=200$ 
as the reference value assuming that $g_*(T)\simeq g_*(T=\mu_{\rm cr})$ 
for temperatures between $m_F$ and $\mu_{\rm cr}$.  
This reference number is thought to be reasonable from the following argument: 
 From the standard model alone we have $g_*|_{\rm SM}=106.75$~\cite{Nakamura:2010zzi}. 
In the one-family model of WTC  
techni-fermions with masses of ${\cal O}({\rm TeV})$ are still in 
the thermal equilibrium  even after its mass generation at $T=\mu_{\rm cr}$ due to the large scale 
hierarchy between $m_F$ and $\mu_{\rm cr}$. 
Techni-gluons are also still massless up to the confinement scale 
typically close/identical to $T=m_F$ to be in the thermal equilibrium. 
Those thermal particles will be relevant participants in the thermal equilibrium, 
which is not to be substantially changed for temperatures between $m_F$ and $\mu_{\rm cr}$. 
Thus the degrees of freedom of relativistic particles in the one-family model can be  
estimated as 
\begin{eqnarray} 
g_*(\mu_{\rm cr}) |_{\rm one-family\,model} &=& \sum_{\rm TG} g_{\rm TG} + \frac{7}{8} \sum_{\rm TF} g_{\rm TF} 
\nonumber \\ 
&=&  (N_{\rm TC}^2-1) \left\{ (2)_{\rm spins} \right\}_{\rm TG}  +  \frac{7}{8} N_{\rm TC} \Bigg[
 \left\{ (2)_{\rm L,R} \times (3)_{\rm color} \times (2)_{\rm isospin} \times (2)_{\rm spins} \right\}_{\rm TQ} 
\nonumber \\ 
&& 
\hspace{40pt} 
+ 
 \left\{ (2)_{\rm L,R} \times (1)_{\rm color} \times (2)_{\rm isospin} \times (2)_{\rm spins} \right\}_{\rm TL}
\Bigg] 
\nonumber \\ 
&=& 
\Bigg\{  
\begin{array}{cc} 
62  & \qquad {\rm for} \qquad N_{\rm TC}=2 \\  
100   & \qquad {\rm for} \qquad  N_{\rm TC}=3 
\end{array}
\,, 
\end{eqnarray} 
 where TG, TQ and TL denote techni-gluon, -quark and -lepton, respectively. 
Adding the standard model contribution leads to $g_{*}(\mu_{\rm cr}) \simeq 176-207$ in total. 

As we have assumed so far, techni-pions can also be in the thermal equilibrium to coexist with techni-fermions and -gluons in the walking 
regime $m_F < T < \mu_{\rm cr}$. The contributions from those scalar particles would shift  
the number of $g_*(\mu_{\rm cr})$ as $g_* \to  g_* + 10$, which corresponds to the shift of about 10\% for $Y_{\rm TD}(T_0)$.

\begin{figure}
\begin{center}
\includegraphics[scale=0.7]{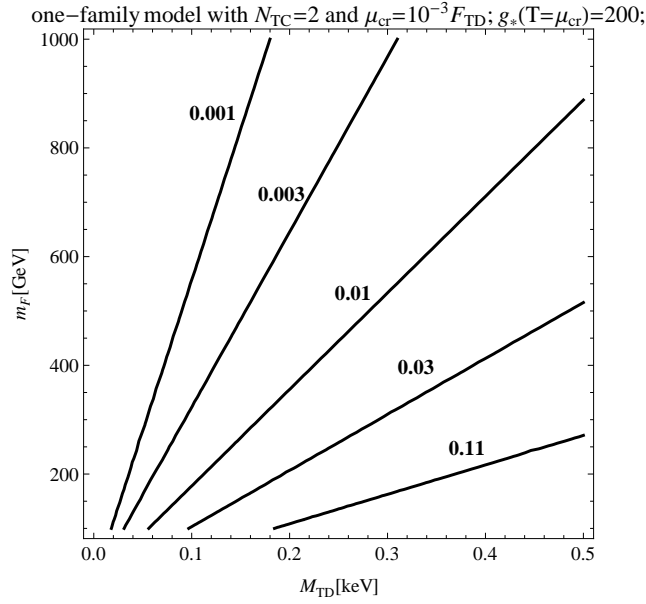}
\end{center}
\caption{The contour plot of the thermally produced relic density of TD $\Omega_{\rm TD}^{\rm tp}h^2$ in the 
$M_{\rm TD}$-$m_F$ plane in the case of one-family model with 
$N_{\rm TC}=2$ and $N_{\rm TF}=8$ together with $\mu_{\rm cr}=10^{-3}F_{\rm TD}$ assumed. 
The labels attached with the contours denote the values of $\Omega_{\rm TD}^{\rm tp}h^2$. 
} 
\label{mF-mTD-tp}
\end{figure}

 From Eq.(\ref{Y:0:onefamily}) 
the relic density is evaluated through the relation $Y_{\rm TD}(T_0)=\frac{n_{\rm TD}(T_0)}{s(T_0)}
=\frac{\Omega_{\rm TD}h^2 \cdot (\rho_{\rm cr}/h^2)}{M_{\rm TD}\cdot s(T_0)}$ to be  
\begin{eqnarray} 
  \Omega^{\rm tp}_{\rm TD} h^2 \Bigg|_{\rm one-family\, model} 
&\simeq&  
\left( \frac{\mu_{\rm cr}}{10^8 {\rm GeV}} \right) 
\left( \frac{M_{\rm TD}}{\rm keV} \right)
\left(\frac{200}{g_*(\mu_{\rm cr})}  \right)^{3/2} 
\left( \frac{10^{11} {\rm GeV}}{F_{\rm TD}} \right)^2 
\nonumber \\ 
&& 
\times 
\Bigg\{  
\begin{array}{cc} 
16 \times 10^{-1}  & \qquad {\rm for} \qquad N_{\rm TC}=2 \\  
1.2 \times 10^{-1} & \qquad {\rm for} \qquad  N_{\rm TC}=3 
\end{array}
\,, \label{relic:tp}
\end{eqnarray}
where use has been made of $s(T_0)=\frac{2\pi^2}{45} g_{*s}(T_0) T_0^3$ with $g_{*s}(T_0)=43/11$, $T_0=2.73$ K$\simeq 2.4 \times 10^{-4}$ eV, 
and $\rho_{\rm cr}/h^2 =0.8 \times 10^{-46}$ GeV$^4$. 
To make a more explicit evaluation 
we may take 
\begin{equation} 
\mu_{\rm cr} \simeq 10^{-3}\Lambda_{\rm TC} \simeq 10^{-3}F_{\rm TD} 
\end{equation} 
inspired by the result of Ref.~\cite{Hashimoto:2010nw}. 
Using the PCDC relation Eq.(\ref{PCDC}) to remove $F_{\rm TD}$ from Eq.(\ref{relic:tp}), 
we rewrite Eq.(\ref{relic:tp}) as  
\begin{eqnarray} 
  \Omega^{\rm tp}_{\rm TD} h^2 \Bigg|_{\rm one-family\, model} 
&\simeq&  
\left(\frac{200}{g_*(\mu_{\rm cr})}  \right)^{3/2} 
\left( \frac{M_{\rm TD}}{\rm keV} \right)^2
\left( \frac{10^{3} {\rm GeV}}{m_F} \right)^2 
\nonumber \\ 
&& 
\times 
\Bigg\{  
\begin{array}{cc} 
5.7 \times 10^{-2}  & \qquad {\rm for} \qquad N_{\rm TC}=2 \\  
2.5 \times 10^{-3} & \qquad {\rm for} \qquad  N_{\rm TC}=3 
\end{array}
\,. \label{relic:tp:2}
\end{eqnarray} 
In Figure~\ref{mF-mTD-tp} we plot the thermally produced relic density 
as a function of $m_F$ and $M_{\rm TD}$ 
in the case of the one-family model with $N_{\rm TC}=2$ and $N_{\rm TF}=8$. 
The values of $M_{\rm TD}$ have been restricted to be in a range $0.01 \lesssim M_{\rm TD} \lesssim 500$ eV which comes from 
some cosmological and astrophysical constraints as will be discussed later.

Figure~\ref{mF-mTD-tp} indicates that the thermally produced relic density becomes 
comparable with the amount of observed dark matter density $\simeq 0.1$ 
if $m_F$ gets smaller than the electroweak scale $\simeq 246$ GeV. 
This is, however, not the case: 
Consider the nonrunning case which yields the mass function of techni-fermion $\Sigma(x)$ ($x=-p^2$) as the solution of Schwinger-Dyson 
equation with the ladder approximation. The $\Sigma(x)$ takes the form $\Sigma(x)=m_F \cdot {}_2F_1(1/2,1/2,2;-x/m_F^2)$ where 
${}_2 F_1$ denotes hyper geometric function. 
The techni-pion decay constant $F_{\pi_T}$ is then calculated using 
the Pagels-Stokar (PS) formula~\cite{Pagels:1979hd} in terms of the dynamical fermion mass $m_F$: 
\begin{eqnarray} 
  F_{\pi_T}^2 &=& 
\frac{N_{\rm TC}}{4 \pi^2} m_F^2 \int_0^{\frac{\Lambda_{\rm TC}^2}{m_F^2} \to \infty} dx \, x  
 \frac{\Sigma^2(x)- \frac{x}{4} \frac{d}{dx} \Sigma^2(x)}{(x + \Sigma^2(x))^2} 
\nonumber \\ 
&\simeq& 
\frac{N_{\rm TC}}{2\pi^2} m_F^2 
\,,  \label{PS}
\end{eqnarray}
for $\gamma_m \simeq 1$. 
On the other hand, the techni-pion decay constant is set by the electroweak scale $\simeq 246$ GeV associated with
 the $W$ and $Z$ boson masses as follows:   
 \begin{eqnarray} 
   F_{\pi_T} \simeq \frac{246{\rm GeV}}{\sqrt{N_{\rm TF}/2}} = 123 \,{\rm GeV}
   \,, \label{Fpit}
 \end{eqnarray}
for the one-family model with $N_{\rm TF}=8$. 
 Using Eqs.(\ref{PS}) and (\ref{Fpit}), we find the techni-fermion mass  of the Pagels-Stokar formular 
\begin{equation} 
  m_F|_{\rm PS} \simeq 386 \, {\rm GeV} 
 \,. \label{PS:mF}
\end{equation} 
Looking at Fig.~\ref{mF-mTD-tp}  we find that the PS value of $m_F$ in Eq.(\ref{PS:mF}) yields 
the thermally produced relic density less than 0.05, which is too small 
to explain the observed dark matter density $\simeq 0.1$. 
It is thus concluded that the thermal population of TD cannot be a main component of 
the dark matter.

Similar estimate is straightforwardly applicable to other 
models of WTC, although the explicit calculation above 
has been specific to the one-family model.

\subsection{Non-thermal production}
\label{NTP}

Analogously to the case of axion dark matter~\cite{Kim:2008hd}, 
the population of TD can be accumulated by ``misalignment"  
of the classical TD field $D$ and the coherent oscillation. 
In this subsection we shall explore this possibility.

\subsubsection{The TD potential}

\begin{figure}[t]

\begin{center} 
\includegraphics[scale=0.7]{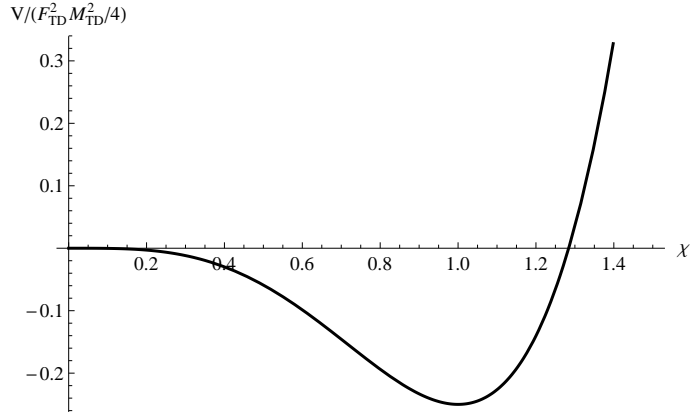}
\end{center}
\caption{The illustration of the TD potential Eq.(\ref{VTD}) 
which has the minimum at $\langle \chi \rangle =1$ with the vacuum energy 
$|V_{\rm min}|=F_{\rm TD}^2M_{\rm TD}^2/16$. } 
\label{potential}
\end{figure}

 Since the TD is pseudo Nambu-Goldstone boson of scale symmetry, 
we work on a nonlinear Lagrangian to address the classical TD dynamics. 
The TD field $D$ is introduced so as to transform nonlinearly under the scale symmetry 
as $\delta D = F_{\rm TD} + x^\nu \partial_\nu D$ with the decay constant $F_{\rm TD}$. 
It is then embedded into the nonlinear base $\chi$ as $\chi = e^{D/F_{\rm TD}}$, 
which transforms with the scale dimension 1: $\delta \chi = (1 + x^\nu \partial_\nu) \chi$. 
The scale invariant TD kinetic term thus takes the form: 
\begin{equation} 
{\cal L}_{\rm kin} 
= \frac{1}{2} F_{\rm TD}^2 (\partial_\mu \chi)^2 
= \frac{1}{2} \chi^2 (\partial_\mu D)^2
\,. \label{D:kin}
\end{equation}
As was introduced in Eq.(\ref{TD:p}), 
in addition, the TD gets the potential~\cite{Migdal:1982jp},  
\begin{equation} 
V_D= \frac{F_{\rm TD}^2 M_{\rm TD}^2}{4} \chi^4 \left( \log \chi - \frac{1}{4} \right) 
 \,, \label{VTD}
\end{equation} 
whose form is depicted in Fig.~\ref{potential} and completely fixed by the scale anomaly Eq.(\ref{PCDC:0}): 
\begin{equation} 
\theta_\mu^\mu = - \delta V_D = - \frac{F_{\rm TD}^2 M_{\rm TD}^2}{4} \chi^4 
 \,. \label{var:VTD}
\end{equation}
which, at the vacuum $\langle \chi \rangle=1$, correctly reproduces the PCDC relation Eq.(\ref{PCDC}).   
The TD potential Eq.(\ref{VTD}) actually takes the same form of that for a QCD dilaton derived 
based on a similar nonlinear realization as discussed in Ref.~\cite{Migdal:1982jp}.

Note that the form of $V_D$ is fairly stable against 
both thermal and radiative corrections: 
thermal corrections are not generated since the TD decouples 
from the thermal equilibrium no sooner than its generation, 
while radiative corrections are almost negligible due to the large suppression by $F_{\rm TD}$.

\subsubsection{The harmonic oscillation}

  From Eqs.(\ref{D:kin}) and (\ref{VTD}), 
we may derive the equation of motion for the zero mode of $\chi$ (i.e. with $\vec{p}=0$) in a flat Friedmann-Robertson-Walker 
metric to find 
\begin{equation} 
  \ddot{\theta} + 3 H(T) \dot{\theta} + 
  \frac{1}{F_{\rm TD}^2} \frac{\partial V_D}{\partial \theta}
= 0
\,, \label{EOM:sigmaD}
\end{equation} 
where the dot denotes derivative with respect to time and  
\begin{eqnarray} 
\theta &\equiv& \chi -1 = \frac{D}{F_{\rm TD}} + \frac{D^2}{2 F_{\rm TD}^2} + \cdots 
\nonumber \\ 
 \frac{\partial V_D}{\partial \theta} 
&=& F_{\rm TD}^2 M_{\rm TD}^2 (1+\theta)^3 \log (1+\theta) 
  \,. 
\end{eqnarray}

\begin{figure}[h]
\begin{center}
\includegraphics[scale=0.7]{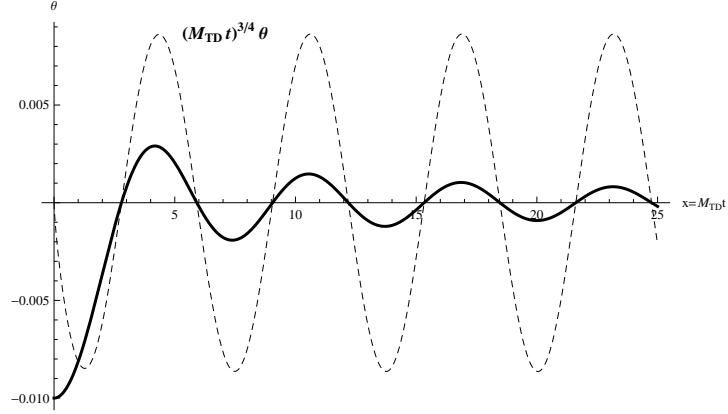}
\end{center}
\caption{
The plots of $\theta$ (solid curve) and $(\MTD\,t)^{3/4} \theta$ (dashed curve)
with respect to $x=M_{\rm TD} t$. 
The boundary condition for $\theta$ is taken to be   
 $\theta(x_i)= - 0.01$ and $\dot{\theta}(x_i)=0$ for $x_i \ll 1$.}
\label{TD-os}
\end{figure}

We shall assume the initial position of $\theta$ in the potential 
in such a way that the $\chi$ immediately starts to move close to the vacuum $\theta =0$, 
i.e., $\theta(t=t_i )\lesssim 1$ with $t_i \times \MTD \ll 1 $ 
and its velocity $\dot{\theta}(t_i)=0$~\footnote{
We have checked that, if the initial position would be apart from the vacuum, 
we would have a too large non-thermal production exceeding the observed dark matter density. 
This is due to the significant anharmonic corrections which become relevant when $\theta \to -1$. 
Other possible initial positions are therefore not to be preferable for TD as a dark matter candidate. }. 
Then the nonlinear terms in the evolution equation Eq.(\ref{EOM:sigmaD}) are not significant 
to be negligible so that 
we have an oscillating solution with a damped harmonic potential around the vacuum at $\theta=0$ 
approximately described by 
$\ddot{\theta} + 3 H \dot{\theta} + M_{\rm TD}^2 \theta  = 0$.  
The TD dynamics is thus to be almost completely adiabatic ($H \ll M_{\rm TD}$) and harmonic oscillation. 
To check this adiabaticity and harmonicity, it is convenient to evaluate 
an adiabatic quantity $I$,   
\begin{eqnarray} 
I &=& R^3(t) \rho_{\rm os}(t)/M_{\rm TD} 
\nonumber \\ 
& \propto& (\MTD\, t)^{3/2} \bar{\theta}^2(t)  
\,, 
\end{eqnarray}
where $R(t)$ is the scale factor, $\bar{\theta}$ is the amplitude of the oscillation
and we used $R(t) \propto \sqrt{t}$ in radiation-dominated epoch. 
Note that $M_{\rm TD}$ is time-independent. 
Since the number density per comoving volume should be preserved 
during the harmonic oscillation, 
the amplitude of $I$ is to be constant as well.  
Solving Eq.(\ref{EOM:sigmaD}) numerically with  
$\theta(t_i)=-0.01$ and $\dot{\theta}(t_i)=0$ taken,  
in Fig.~\ref{TD-os} we plot $(\MTD\, t)^{3/4} \theta(t) $ (dashed
  curve) together with $\theta$ (solid curve) as a function of $x=M_{\rm TD} t$.  
Noting that the amplitude of $(\MTD\, t)^{3/4} \theta(t) $ is proportional to $\sqrt{I}$, 
we see that Fig.~\ref{TD-os} indeed shows that the harmonic oscillation starts no later than  
the $\sigma_D$ starts moving to roll down to the vacuum in the potential.

Thus the adiabatic harmonic oscillation well approximates the TD dynamics when 
$\theta_i \ll 1$ so that the time $T_{\rm os}$ when the oscillation starts 
is estimated through $3H(T_{\rm os})=\MTD$ to be 
\begin{eqnarray} 
T_{\rm os} &\simeq& 
4.2 \times 10^5 \, {\rm GeV}  \left( \frac{M_{\rm TD}}{{\rm keV}} \right)^{1/2}
\left( \frac{200}{g_*(T_{\rm os})} \right)^{1/4} 
\,. 
\end{eqnarray}  
Note that $T_{\rm os}$ is much smaller than the scale of TD generation, $\mu_{\rm cr} \simeq 10^8$ GeV, 
and hence the TD does not start the oscillation 
until $T=T_{\rm os}$ though it was generated at the far past time $T = \mu_{\rm cr}$.

\subsubsection{The estimate of the non-thermally produced TD relic density}

\begin{figure}
\begin{center}
\includegraphics[scale=0.6]{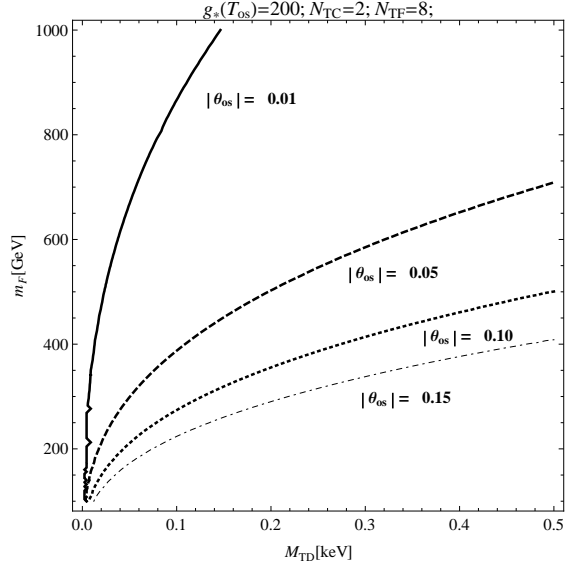}
\end{center}
\caption{The contour plot in the $M_{\rm TD}$-$m_F$ plane 
derived by requiring $\Omega_{\rm TD}^{\rm ntp}h^2 \le 0.11$ 
in the case of one-family model with $N_{\rm TC}=2$ and $N_{\rm TF}=8$. 
In the plot the curves have been drawn by changing 
the misalignment parameter $|\theta_{\rm os}|$ to be 0.01 (solid), 0.05 (dashed), 0.10 (dotted) and 0.15 (dot-dashed) 
as the reference values. 
The regions above the curves are excluded.} 
\label{NTP-contour}
\end{figure}

The conservation of number density per comoving volume during the oscillation 
leads to 
\begin{equation} 
\rho_{\rm TD}(T_0) 
= \rho_{\rm TD}(T_{\rm os}) \cdot 
\frac{s(T_0)}{s(T_{\rm os})} 
\,.  \label{scaling:rho}
\end{equation} 
The TD energy density at $T=T_{\rm os}$, $\rho_{\rm TD}(T_{\rm os})$, is evaluated as 
\begin{equation} 
\rho_{\rm TD}(T_{\rm os}) =  |V(T_{\rm os}) -  V_{\rm min}|    
\,. 
\end{equation} 
We expand $V(T_{\rm os})$ around the vacuum $\theta=0$:  
\begin{eqnarray} 
V(T_{\rm os}) &=& \frac{F_{\rm TD}^2 M_{\rm TD}^2}{4} (1 + \theta_{\rm os})^4 (\log (1 + \theta_{\rm os}) - 1/4) 
\nonumber \\ 
&\simeq& 
-\frac{F_{\rm TD}^2 M_{\rm TD}^2}{16} \left( 1 - 8 \theta_{\rm os}^2 + {\cal O}(\theta_{\rm os}^3) \right)
\,, 
\end{eqnarray} 
such that 
\begin{eqnarray} 
\rho_{\rm TD}(T_{\rm os}) &=& 
\frac{F_{\rm TD}^2 M_{\rm TD}^2}{2} \theta_{\rm os}^2 
\nonumber \\ 
&=& 
\frac{8 N_{\rm TC}N_{\rm TF}}{\pi^4} m_F^4  \theta_{\rm os}^2 
\,,   \label{rho:Tos}
\end{eqnarray} 
where in the last line we have used the PCDC relation Eq.(\ref{PCDC}). 
Using Eqs.(\ref{scaling:rho}) and (\ref{rho:Tos})  
we thus calculate the non-thermally produced relic density to get   
\begin{eqnarray} 
\Omega_{\rm TD}^{\rm ntp}h^2 
&\simeq&  
11
\times 
\left( \frac{\theta_{\rm os}}{0.05} \right)^2
 \left(\frac{200}{g_*(T_{\rm os})} \right)  \left(\frac{m_F}{10^3 {\rm GeV}} \right)^4
  \left(\frac{10^5 {\rm GeV}}{T_{\rm os}} \right)^3, 
\label{non-ther-TD-relic}
\end{eqnarray} 
 for the one-family model with $N_{\rm TC}=2$ and $N_{\rm TF}=8$.  
Figure~\ref{NTP-contour} shows some contours in the $M_{\rm TD}$-$m_F$ plane 
obtained by requiring $\Omega_{\rm TD}^{\rm ntp}h^2 \le 0.11$, 
where the regions above the curves are excluded. 
Here we have taken $g_*(T_{\rm os})=200$ and restricted $M_{\rm TD}$ to a region 
$0.01 \lesssim M_{\rm TD} \lesssim 500$ eV as was done in Fig.~\ref{mF-mTD-tp}. 
As the reference values the misalignment parameter $\theta_{\rm os}$ has been chosen to be 0.01, 0.05, 0.10, and 0.15. 
  Figure~\ref{NTP-contour} 
implies that the misalignment with $|\theta_{\rm os}| \lesssim 0.15$ is preferable 
for the PS value of $m_F$, $m_F \simeq 386$ GeV in Eq.(\ref{PS:mF}). 
It is thus concluded that the non-thermal TD can be produced 
enough to explain the observed dark matter density.

\section{Cosmological and astrophysical constraints} 
\label{constraint}
In this section we shall address several constraints on our light decoupled 
TD coming from cosmological and astrophysical observations at present. 

\subsection{TD relic density for dark matter}

\begin{figure}
\begin{center}
\includegraphics[scale=0.6]{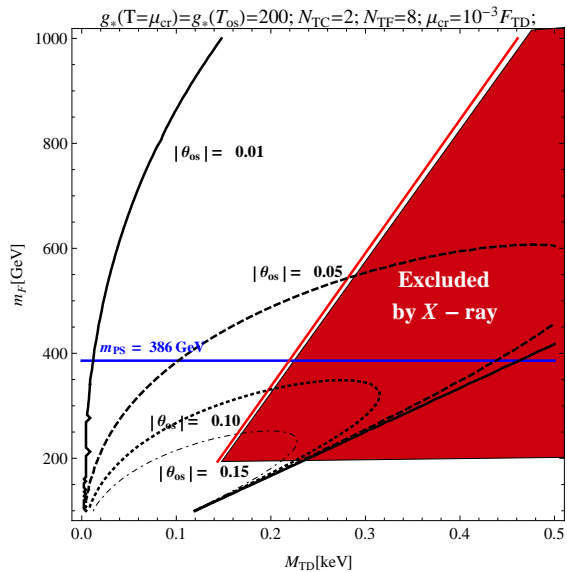}
\end{center}
\caption{The contour plot of the  total relic density of TD from thermal and non-thermal production in  
the $M_{\rm TD}$-$m_F$ plane satisfying $\Omega_{\rm TD}^{\rm tot} h^2 = 0.11$ 
for the one-family model with $N_{\rm TC}=2$ and $N_{\rm TF}=8$. 
In the figure the solid, dashed, dotted and dot-dashed curves correspond to 
choice of the misalignment parameter $|\theta_{\rm os}|=$ 
0.01, 0.05, 0.10 and 0.15, respectively.  
The domain inside the region surrounded by each curve is 
allowed and outside is 
excluded so that the TD relic density exceeds the presently observed dark matter 
density. The X-ray constraint has been incorporated in the plot by the red region. 
The PS value of $m_F$ is shown as the horizontal blue line. }
\label{tot-contour}
\end{figure}

Combining Eqs.(\ref{relic:tp:2}) and (\ref{non-ther-TD-relic}) we now evaluate 
the total TD relic density: 
\begin{equation} 
\Omega_{\rm TD}^{\rm tot} h^2 = \Omega_{\rm TD}^{\rm tp} h^2 + \Omega_{\rm TD}^{\rm ntp} h^2
\,. 
\end{equation}  
For the total dark matter relic density, we know 
$0.109 < \Omega_{DM} h^2   <0.113$ for the $3\sigma$ range observed at the WMAP-7~\cite{Komatsu:2010fb}.  
Here we take $\Omega_{DM} h^2 = 0.11$ as the reference value for simplicity.  
Figure~\ref{tot-contour} shows the contours 
of $\Omega_{\rm TD}^{\rm tot} h^2 = 0.11$  in the $M_{\rm TD}$-$m_F$ plane
for the one-family model with $N_{\rm TC}=2$ and $N_{\rm TF}=8$. 
Here we used  $|\theta_{\rm os}|=$ 
0.01, 0.05, 0.10 and 0.15 corresponding to the solid, dashed, dotted and dot-dashed 
lines  respectively. 
The region outside of the contours 
is excluded since the TD relic density exceeds the presently observed dark matter 
density.  Therefore for given $m_F$, the lower bound on $M_{\rm TD} $ comes from the excessive   non-thermal production of TD  (See also Fig.~\ref{NTP-contour}) and the upper bound comes from  
the excessive thermal production.

As will turn out below, 
the X-ray observation severely constrains $M_{\rm TD}$ provided  
the TD fully saturate the present dark matter density. 
The X-ray constraint has also been incorporated into the plot in Fig.~\ref{tot-contour} by the red region. 

\begin{figure}[h]
\begin{center}
\includegraphics[scale=0.6]{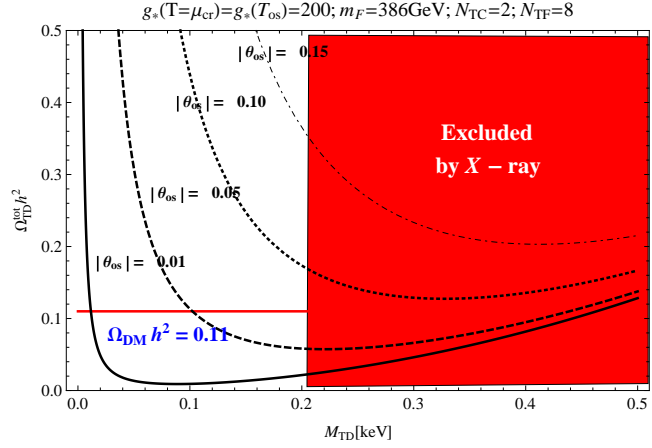}
\end{center}
\caption{The total TD relic density 
as a function of $M_{\rm TD}$ in the case of one-family model with 
$N_{\rm TC}=2$ and $N_{\rm TF}=8$. 
The value of $m_F$ is fixed as $m_F=386$ GeV using the Pagels-Stokar formula. 
The solid, dashed, dotted and dot-dashed curves correspond to choice of the misalignment parameter $\theta_{\rm os}$, 
$|\theta_{\rm os}|=0.01, 0.05, 0.10$ and 0.15, respectively. 
The horizontal straight line denotes the presently observed dark matter density. 
} 
\label{tot-omega-mTD-w-mPS}
\end{figure}

In Fig.~\ref{tot-omega-mTD-w-mPS} 
we also make a plot of the total TD relic density as a function of $M_{\rm TD}$ 
fixing $m_F$ to the PS value in Eq.(\ref{PS:mF}), in the same fashion as in Fig.~\ref{tot-contour}. 
 Figure~\ref{tot-omega-mTD-w-mPS} tells us that, if 
the observed dark matter density is fully filled with TD to be consistent with the X-ray constraint,  
the TD mass is predicted to be       
\begin{equation}  
M_{\rm TD} \simeq  100 \, {\rm eV}
\,, \label{010}
\end{equation}   
for $|\theta_{\rm os}|=0.05$ and 
\begin{equation}  
M_{\rm TD} \simeq 10 {\rm eV}
\,, \label{005}
\end{equation}   
for $|\theta_{\rm os}|=0.01$. 
We thus conclude that, for $|\theta_{\rm os} |\lesssim 0.05$,  
the light decoupled TD can indeed be a dark matter candidate 
consistently with astrophysical and cosmological constraints.

\subsection{Test of the gravitational inverse-square law}

\begin{figure}
\begin{center}
\includegraphics[scale=0.5]{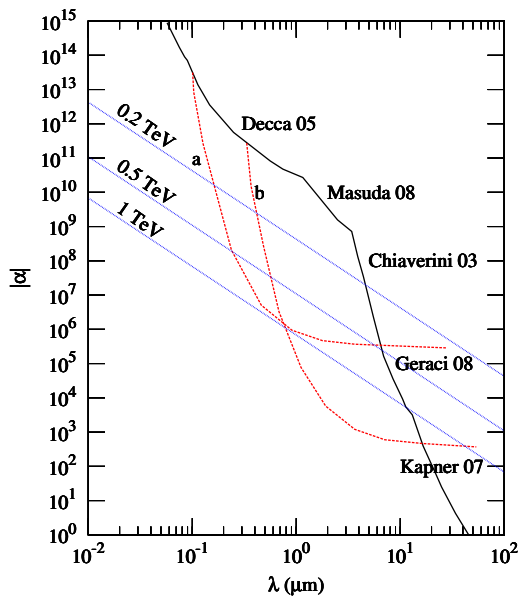}
\end{center}
\caption{The experimental constraints (black solid line) and the projected
  improved search (two red lines) on new forces from Yukawa-type potential below
100 $\mu m$. The region above the lines are disallowed. Experimental data are adapted from a figure in
Ref.~\cite{Geraci:2010ft}. The theoretical predictions from TD are shown for
$m_F=0.2,0.5$ and $1\tev$ with blue lines. }
\label{Geraci_TD}
\end{figure}

The search for violations of the gravitational inverse-square law can
constrain short range forces from the exchange of scalar or vector particles.
Corrections to Newtonian gravity at short range are generally parameterized 
in terms of a Yukawa-type potential
\dis{
V=-\frac{G_N m_1m_2}{r}\left[1+\alpha e^{-r/\lambda}  \right],
\label{Yukawa-type}
}
where $m_1$ and $m_2$ are masses of nucleons interacting at distance $r$, $\alpha$
denotes the overall strength of the Yukawa correction relative to gravity, and $\lambda$ is the
effective range of the Yukawa interaction. 
The current experimental upper limits (black solid line) and the projected upper limits (two red lines) on $|\alpha|$ are 
shown in the figure \ref{Geraci_TD} for $\lambda < 100 \mu m$~\cite{Geraci:2010ft}. 

  The coupling of TD to a nucleon $\psi_N$ can be expressed by 
\dis{
\beta_N\frac{2m_N}{\FTD} \sigma_D \bar{\psi_N} \psi_N, 
\label{TD:yukawa}
}
where $m_N$ is the nucleon mass and $\beta_N$ an overall coefficient associated with low-energy hadron physics 
which is assumed to be ${\mathcal O}(1)$. 
  From the Yukawa interaction in Eq.(\ref{TD:yukawa}) we read off the size of $\alpha$  in Eq.(\ref{Yukawa-type}) 
  relative to 
the gravitational interaction as 
\dis{
\alpha=8\beta_N^2\bfrac{\MP^2}{\FTD^2}.
\label{alpha:TD}
}
While the Compton wavelength of TD is given as 
\dis{
\lambda_{TD}=\frac{\hbar}{\MTD\, c}= 0.197\, \mu m\bfrac{\ev}{\MTD}.
\label{lambda:TD}}
  Using Eqs.(\ref{alpha:TD}) and (\ref{lambda:TD}) together with the PCDC relation Eq.(\ref{PCDC}), 
in Fig.~\ref{Geraci_TD} we plot $\alpha$ as a function of $\lambda$ 
  in the case of the one-family model with $N_{\rm TC}=2$ and $N_{\rm TF}=8$ 
taking $m_F=0.2$, 0.5, $1 \tev$, in comparison with the exclusion curves from the experimental data. 
  From this figure we see that the present experiment gives an upper bound on the Compton length of TD
\dis{
\lambda_{TD} \lesssim  5 - 20 \,\mu m ,
}
which is, from Eq.(\ref{lambda:TD}), rephrased in terms of the lower bound on the TD mass, 
\begin{equation} 
\MTD \gtrsim 0.01-0.04\, \ev.
\end{equation}


\subsection{X-ray background}

\begin{figure}
\begin{center}
\includegraphics[scale=0.5]{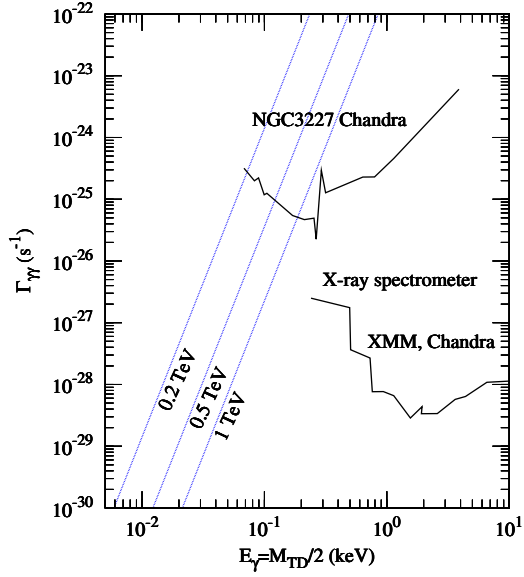}
\end{center}
\caption{Experimental constraints (black solid line) from X-ray background. 
The regions above the lines are disallowed.
Experimental data are adapted from a figure in
Ref.~\cite{Bazzocchi:2008fh}. The theoretical predictions from TD are shown for
$m_F=0.2,0.5$ and $1\tev$ with blue lines. }
\label{XRB_TD}
\end{figure}

The light TD dark matter decays dominantly into two photons with the energy
$E_\gamma=\MTD/2$. Thus the TD in the keV range can emit a mono-energetic 
X-ray which can be detected in the X-ray observation.
The photons produced in the TD dark matter decay will show up in the diffuse 
X-ray background and also in the X-ray emission from the dark-matter-dominated 
objects. Thus those observation can constrain the decay rate of TD into
photons  model independent way.
We use the upper limit on the decay rate from the analysis of NGC3227 
using the high
resolution camera (HRC) carried in the X-ray observatory Chandra, and other
X-ray observations used in~\cite{Bazzocchi:2008fh}. 
In figure~\ref{XRB_TD} we show the decay rate of TD Eq.(\ref{lifetime}) 
 for the one-family model 
quoted in Fig.~\ref{tot-contour} in the range of $m_F=0.2 \tev - 1 \tev$. 
The observational constraints on the decay rate are taken 
from~\cite{Bazzocchi:2008fh}. 
It turns out that the X-ray observation excludes the following region in the $(M_{\rm TD},m_F)$ plane 
(for $m_F \gtrsim 0.2$ TeV), 
\begin{equation} 
 m_F \, [{\rm TeV}] \le 0.249 \, M_{\rm TD}\, [{\rm keV}] - 0.148 
 \,, 
\end{equation}
which gives the upper bound on $M_{\rm TD}$,  
$\MTD \lesssim (140, 260, 460)
\ev$  for $m_F=(0.2,0.5,1)\tev$, respectively.

\subsection{Neutron star cooling}

Since TD couples to nucleons, 
the TD decay constant $\FTD$ may be constrained by energy loss in stars 
through the TD production out of stars similarly to the case of axion~\cite{Kim:2008hd}. 
The most severe constraint would then come from neutron star cooling to 
give a lower bound on $F_{\rm TD}$, $\FTD \gtrsim 10^9\gev$. 
One can see, however, that this constraint is well satisfied if the lifetime of TD is long enough 
for TD to be a dark matter 
as in~\eq{lifetime}: indeed, it just places a weaker upper bound on the TD mass, $M_{\rm TD}\lesssim 1\mev$, 
in contrast to the case of axion, 
which is due to the discrepancy between their energy densities, 
namely, $F_{\rm TD}^2M_{\rm TD}^2 \sim m_F^4$ and $f_a^2 m_a^2 \sim \Lambda_{\rm QCD}^4$.

\section{TD dark matter detection experiments} 
\label{detection}

Since the light decoupled TD has the extremely large decay constant $F_{\rm TD} \gtrsim 10^{12}$ GeV 
and its lifetime has to be $\lesssim 10^{-25}$s to be consistent with X-ray observation (See Fig.~\ref{XRB_TD}), 
it is unlikely that the TD dark matter can be detected at collider experiments as in the case of invisible axions.   
However, detection experiments through cosmological and astrophysical sources can be accessible due to the TD-photon 
coupling analogously to the axion case~\cite{Kim:2008hd}.  
In this section we shall give several comments on such a detection potential.

The experiments on new forces through modified Newtonian gravity 
may give a possibility to detect a light TD. 
A new experiment proposed by Geraci et al.~\cite{Geraci:2010ft} using optically
trapped and cooled dielectric microsphere can be used to establish the
existence of TD with the mass around $0.1 \sim 1 \ev$ as shown with red
(solid)
line in Fig.~\ref{Geraci_TD}.
The TD dark matter may actually be quite difficult to detect 
since the mass is constrained to be larger than around $0.1\kev$ and 
hence it generates an extremely short ranged fifth force.

However, a monoenergetic X-ray signal produced from the TD-photon conversion in the sky  
would provide a possible chance to detect the existence of TD indirectly.

The TD detection experiments can be performed associated with the TD-photon conversion, 
in a way similar to the axion case~\cite{Kim:2008hd}:  
From \eq{gaugeint:anomalous} the TD coupling to two photons are read off as 
\dis{
{\mathcal L}_{D \gamma \gamma}=-\alpha_1DF_{\mu\nu}F^{\mu\nu}=2\alpha_1 D
(\vec{E}^2-\vec{B}^2), \label{Dgg}
}
where 
\dis{
\alpha_1=
\frac{(3-\gamma_m)\beta(e)}{2 e}\frac{1}{\FTD}=\frac{(3-\gamma_m)\alpha_{\rm
    EM}}{6 \pi}\frac{|{\cal C}-5/2|^2}{\FTD} \qquad 
}
and ${\cal C} \equiv \sum_{F} N_{\rm TC} N_c^{F} Q_{F}^2 $. 
This coupling is similar to that of axion except the essential fact that axion is a pseudo scalar and hence 
couples to photon via $\vec{E}\cdot\vec{B}$. 
One possible example of TD detection experiments is the following: 
Cosmic TDs left over from the big bang may be detected using microwave cavity
haloscopes similar to the axion case~\cite{SikivieDetection}.
The TDs drift through the microwave cavity in a strong static magnetic
field and resonantly convert to microwave photons according to \eq{Dgg}.
Since TD is a scalar particle, they will convert their energy into a TE mode (magnetic wave)
in the cavity. The conversion power $P$ is then given by~\cite{Cho:2007cy}
\dis{
P=\bfrac{4\alpha_1^2}{\pi^2}\rho B_0^2L_xV,\label{power}
}
where $\rho$ is the local TD energy density, $B_0$ is the magnetic strength, 
$V$ is the volume of the cavity and $L_x$ is the size of $x$ direction.
For the experimental setup with $B_0=10\, {\rm T}, $ $L_x=1\, {\rm m}$, $V=1
\,{\rm m}^3$ and the local halo density $\rho_{\rm halo}\simeq0.3\, {\rm GeV}/{\rm
  cm}^3$, \eq{power} then reads
\dis{
P=2.5 \times 10^{-22}\, {\rm Watt}\,
\bfrac{C}{17/6}\bfrac{10^{12}\gev}{\FTD}\bfrac{\rho}{\rho_{\rm halo}}.
} 
This power is comparable to that of axion and it might be possible
to detect cosmological TDs in the cavity experiment with appropriate
experimental device.

\section{Summary} 
\label{summary}

To summarize, this paper explored  a new dark matter candidate, light decoupled TD,   
arising as a pseudo Nambu Goldstone boson in 
the almost conformal dynamics of WTC possessing   
the approximate scale invariance. 
The extremely WTC allows TD to be extremely light 
with mass much smaller than the techni-fermion mass scale and its decay constant comparable with 
the cutoff scale of WTC. 

The light decoupled TD was studied in detail 
explicitly by discussing cosmological productions of TD, thermal and non-thermal productions, 
in the early Universe. 
The thermal population is governed by single TD production processes involving vertices 
breaking the scale symmetry (See Figs.~\ref{leading-graphs-TD} and \ref{td-production}), 
while the non-thermal population is accumulated via 
harmonic and coherent oscillations of classical TD field similarly to cold axion dark matter case (Figs.~\ref{NTP-contour}). 
It was shown that 
the non-thermal population is dominant and large enough to explain the abundance of presently 
observed dark matter (Fig.\ref{NTP-contour}). 
On the other hand, the thermal TD population tends to be highly suppressed due to the large TD decay constant (Fig.~\ref{mF-mTD-tp}).     
It was also clarified that several cosmological and astrophysical observations constrain 
the mass $M_{\rm TD}$ to be in a range, $0.01 {\rm eV} \lesssim M_{\rm TD} \lesssim 500$ eV (See Figs.~\ref{Geraci_TD} and \ref{XRB_TD}).   
The combined result of two cosmological productions including those constraints 
drew a conclusion that the light decoupled TD can be a good candidate for dark matter with the mass around a few hundreds of eV for 
a typical model of WTC (Fig.~\ref{tot-contour}). 
Possible designated experiments to detect the TD dark matter were also explored.

\section*{Acknowledgments}
We would like to thank S. Jung, J.~E. Kim, B. Kyae, F. Sannino, and K. Yamawaki for useful comments. 
This work was supported by the Korea Research 
Foundation Grant funded by the Korean Government
KRF-2008-341-C00008 (D.\,K.\,H.) and No. 2011-0011083 (K.\,Y.\,C.).  
KYC acknowledges the Max Planck Society (MPG),
  the Korea Ministry of Education, Science and Technology (MEST),
  Gyeongsangbuk-Do and Pohang City for the support of the Independent
  Junior Research Group at the Asia Pacific Center for Theoretical
  Physics (APCTP).
  S.M. was supported in part by 
the JSPS Grant-in-Aid for Scientific Research (S) \#22224003.
Part of this work was done during APCTP focus program on Holography at LHC.

\end{document}